\documentclass[prb,reprint,floatfix,groupedaddress,superscriptaddress,showpacs]{revtex4-1}

\usepackage{amsmath}
\usepackage{amssymb}
\usepackage{amsfonts}
\usepackage{longtable}
\usepackage{bbm}
\usepackage{color}  
\usepackage{graphicx}
\usepackage{dcolumn}
\usepackage{bm}
\usepackage{epsfig}
\usepackage{natbib}
\usepackage{multirow}
\usepackage{hyperref}

\newcommand\T{\rule{0pt}{2.6ex}}
\newcommand\B{\rule[-1.2ex]{0pt}{0pt}}

\newcolumntype{.}{D{.}{.}{3.6}}

\begin{document}

\title{120$^{\circ}$ Helical Magnetic Order in the Distorted Triangular Antiferromagnet $\alpha$-CaCr$_2$O$_4$}

\author{S. \surname{Toth}}
\email{sandor.toth@helmholtz-berlin.de}
\affiliation{Helmholtz Zentrum Berlin f\"{u}r Materialien und Energie, Hahn-Meitner-Platz 1, D-14109 Berlin, Germany}
\affiliation{Institut f\"{u}r Festk\"{o}rperphysik, Technische Universit\"{a}t Berlin, Hardenbergstr.\ 36, D-10623 Berlin, Germany}

\author{B. Lake}
\affiliation{Helmholtz Zentrum Berlin f\"{u}r Materialien und Energie, Hahn-Meitner-Platz 1, D-14109 Berlin, Germany}
\affiliation{Institut f\"{u}r Festk\"{o}rperphysik, Technische Universit\"{a}t Berlin, Hardenbergstr.\ 36, D-10623 Berlin, Germany}
\date{\textrm{\today}}

\author{S. A. J. Kimber}
\affiliation{ESRF, 6 Rue Jules Horowitz BP 220, F-38043 Grenoble Cedex 9, France}
\affiliation{Helmholtz Zentrum Berlin f\"{u}r Materialien und Energie, Hahn-Meitner-Platz 1, D-14109 Berlin, Germany}

\author{O. Pieper}
\affiliation{Helmholtz Zentrum Berlin f\"{u}r Materialien und Energie, Hahn-Meitner-Platz 1, D-14109 Berlin, Germany}

\author{M. Reehuis}
\affiliation{Helmholtz Zentrum Berlin f\"{u}r Materialien und Energie, Hahn-Meitner-Platz 1, D-14109 Berlin, Germany}

\author{A. T. M. N. Islam}
\affiliation{Helmholtz Zentrum Berlin f\"{u}r Materialien und Energie, Hahn-Meitner-Platz 1, D-14109 Berlin, Germany}

\author{O. Zaharko}
\affiliation{Laboratory for Neutron Scattering, PSI, CH-5232 Villigen, Switzerland}

\author{C. Ritter}
\affiliation{Institut Laue-Langevin, BP 156, F-38042 Grenoble Cedex 9, France}

\author{A. H. Hill}
\affiliation{ESRF, 6 Rue Jules Horowitz BP 220, F-38043 Grenoble Cedex 9, France}

\author{H. Ryll}
\affiliation{Helmholtz Zentrum Berlin f\"{u}r Materialien und Energie, Hahn-Meitner-Platz 1, D-14109 Berlin, Germany}

\author{K. Kiefer}
\affiliation{Helmholtz Zentrum Berlin f\"{u}r Materialien und Energie, Hahn-Meitner-Platz 1, D-14109 Berlin, Germany}

\author{D. N. Argyriou}
\affiliation{Helmholtz Zentrum Berlin f\"{u}r Materialien und Energie, Hahn-Meitner-Platz 1, D-14109 Berlin, Germany}

\author{A. J. Williams}
\affiliation{Department of Chemistry, Princeton University, Princeton, New Jersey 08544, USA}


\begin{abstract}
$\alpha$-CaCr$_2$O$_4$ is a distorted triangular antiferromagnet. The magnetic Cr$^{3+}$ ions which have spin-3/2 and interact with their nearest neighbors via Heisenberg direct exchange interactions, develop long-range magnetic order below $T_N=42.6$ K. Powder and single-crystal neutron diffraction reveal a helical magnetic structure with ordering wavevector ${\bf k}=(0, \sim1/3, 0)$ and angles close to $120^\circ$ between neighboring spins. Spherical neutron polarimetry unambiguously proves that the spins lie in the $ac$ plane perpendicular to ${\bf k}$. The magnetic structure is therefore that expected for an ideal triangular antiferromagnet where all nearest neighbor interactions are equal, in spite of the fact that $\alpha$-CaCr$_2$O$_4$ is distorted with two inequivalent Cr$^{3+}$ ions and four different nearest neighbor interactions. By simulating the magnetic order as a function of these four interactions it is found that the special pattern of interactions in $\alpha$-CaCr$_2$O$_4$ stabilizes $120^\circ$ helical order for a large range of exchange interactions.
\end{abstract}

\maketitle

\section{Introduction}
Frustrated magnets, characterized by competing magnetic interactions, continue to generate much research interest. In these systems it is not possible to satisfy all magnetic interactions simultaneously, as a result the ground state can be highly degenerate leading to exotic physical states e.g. spin liquid behaviour and chiral order. The simplest frustrated system is the triangular lattice antiferromagnet where all magnetic interactions between nearest neighbors ($J_\text{nn}$) are equal. In their pioneering work, Anderson and Fazekas suggested that the ground state is a spin liquid with no long-range magnetic order. \cite{Anderson73, Fazekas74} However, recent theoretical work implies that at $T=0$ it develops  long-range order \cite{Capriotti1999} with a helical structure, where the spin moments on nearest neighbors point $120^\circ$ with respect to each other. Interactions between the planes can stabilize this ground state at finite temperatures.

Among real triangular lattice materials the $120^\circ$ structure has been found in CsFe(SO$_4$)$_2$, RbFe(SO$_4$)$_2$, \cite{Serrano1999} VX$_2$ (X=Cl, Br, I), \cite{Hirakawa1983} CuCrO$_2$, \cite{Kadowaki1990} Ag$_2$NiO$_2$, \cite{Nozaki2008} and RbFe(MoO$_4$)$_2$. \cite{Svistov2003} Unfortunately, these compounds only exist in polycrystalline form or as small single-crystals limiting the possibilities of experimental investigation. Additional terms in the spin Hamiltonian can favour different magnetic structures or even destroy long range order. For example single-ion anisotropy favors collinear order (CuFeO$_2$, \cite{Mitsuda1991} $\alpha$-NaMnO$_2$ \cite{Giot2007}); frustrated interlayer interactions suppress order (AgCrO$_2$, \cite{oohara1994} NaCrO$_2$ \cite{hsieh2008}) and strong exchange striction drive the system away from $120^\circ$ helical magnetism ($\alpha$-NaMnO$_2$, \cite{Zorko2008} CuCrS$_2$ \cite{Winterberger1987, Rasch2009}).

Departures from ideal triangular crystal symmetry typically lead to spatially anisotropic exchange interactions sometimes accompanied by orbital ordering and resulting in the frustration being lifted. In most cases the distortion reduces the dimensionality, so that one of the three nearest neighbor exchange interactions ($J_\text{nn1}$) is stronger than the other two ($J_\text{nn2}$), and together the intraplanar interactions produce antiferromagnetic chains with frustrated interchain interactions. The magnetic structure is often helical but where the angle between nearest neighbours along the chain takes a value between $120^{\circ}$ and $180^{\circ}$.

Examples of such compounds are Cs$_2$CuCl$_4$ where $J_\text{nn2}/J_\text{nn1}=0.33$ and the angle between intrachain neighbours is $170^{\circ}$, \cite{coldea2001} while for $\alpha$-NaMnO$_2$ the presence of single ion anisotropy along with the ratio $J_\text{nn2}/J_\text{nn1}=0.44$ gives rise to collinear antiferromagnetic order. \cite{Giot2007,Zorko2008} A third example is CuCrO$_2$, a multiferroic compound where a small lattice distortion accompanies the transition to long-range magnetic order. This along with substantial next nearest neighbour and frustrated interplane interactions, results in angles between nearest neighbors of 118$^{\circ}$ and 120$^{\circ}$. \cite{Poienar2010,Kimura2009}
\begin{figure}[!htb]
  \centering
  \includegraphics[width= 86 mm]{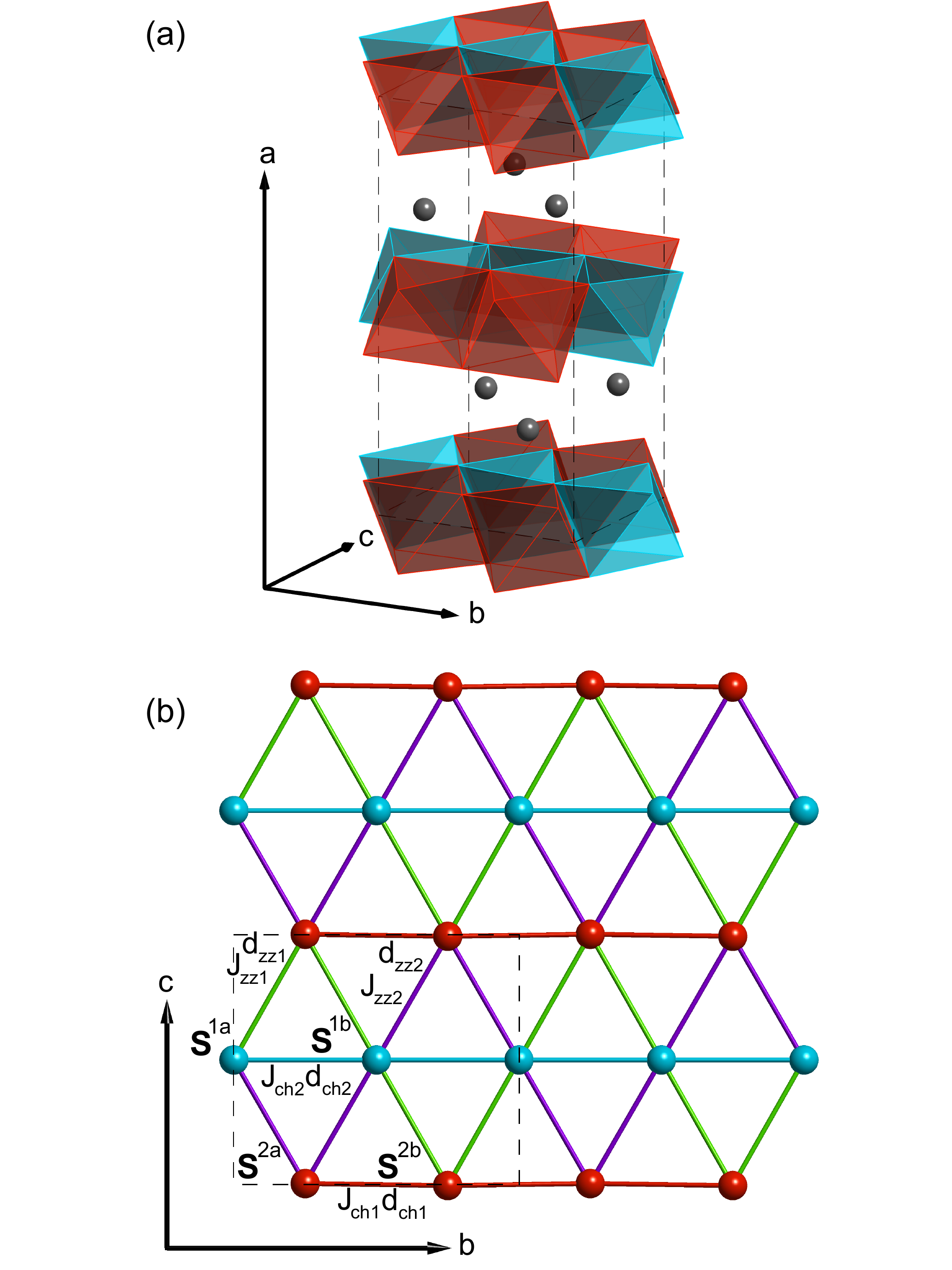}
  \caption{Crystal structure and magnetic interactions of $\alpha$-CaCr$_2$O$_4$.
    (a) Crystal structure showing the triangular layers formed from two inequivalent CrO$_6$ octahedra colored blue for Cr(1)O$_6$ and red for for Cr(2)O$_6$. The layers are separated by Ca$^{2+}$ ions. 
    (b) A single triangular layer showing only the Cr$^{3+}$ ions. The four inequivalent nearest-neighbour intralayer Cr$^{3+}$--Cr$^{3+}$ distances determined from the E9 powder data  at 2.1 K (see Table \ref{tab:rlu}): red $d_\text{ch1}=2.882$ \AA; blue $d_\text{ch2}=2.932$ \AA; green $d_\text{zz1}=2.914$ \AA; purple $d_\text{zz2}=2.910$ \AA. The corresponding exchange interactions are $J_\text{ch1}$, $J_\text{ch2}$, $J_\text{zz1}$ and $J_\text{zz2}$, (see Discussion section). The dashed lines indicate the boundaries of the structural unit cell.}
  \label{fig:dist}
\end{figure}

In this paper we will investigate $\alpha$-CaCr$_2$O$_4$ which is a triangular lattice antiferromagnet belonging to the delafossite arisotype. \cite{Pausch1974} Although this compound is also distorted from ideal triangular symmetry (orthorhombic space group $Pmmn$), the distortion does not lower the dimensionality and is of a different type to that explored previously. Figure \ref{fig:dist} shows the crystal structure. There are two inequivalent Cr$^{3+}$ ions (spin-3/2) which together form distorted triangular layers in the $bc$ plane, that are stacked along $a$. The Cr$^{3+}$ ions are in an octahedral environment and the three $d$ electrons fill the three $t_\text{2g}$ orbitals resulting in quenched orbital moment and negligible anisotropy. Neighboring CrO$_6$ octahedra are edge-sharing leading to direct overlap of the $t_\text{2g}$ orbitals and this along with the short Cr$^{3+}$--Cr$^{3+}$ distances ($\sim 3$ \AA) result in direct antiferromagnetic exchange interactions. Superexchange via oxygen is suppressed because the Cr--O--Cr angles are $\sim 90^\circ$. There are a total of four different nearest neighbor distances giving four inequivalent exchange interactions which together form the complex pattern shown in Fig.\ \ref{fig:dist}(b). Interplane distances are much larger ($\sim 5.5$ \AA) and only weak superexchange interactions are expected between the layers. Recently the magnetic properties of $\alpha$-CaCr$_2$O$_4$ were investigated by Chapon {\em et al}. \cite{Chapon2011} Using powder samples they show that long range magnetic order develops below $T_N$=43 K with ordering wavevector ${\bf k}=(0,0.3317(2),0)$ and spins lying in the $ac$ plane. The magnetic structure is indirectly inferred to be helical rather than sinusoidal so as to constrain the size of chromium moment within physical limits, and the angles between nearest neighbor spins are almost $120^\circ$. Using symmetry arguments it is shown that this compound is not ferroelectric although quadratic magneto-electric effects are possible.

We have grown the first single-crystals of $\alpha$-CaCr$_2$O$_4$ and in this paper we investigate the magnetic properties using heat capacity, DC magnetic susceptibility, neutron and X-ray powder diffraction, neutron single-crystal diffraction and spherical neutron polarimetry. We show that despite the fact that the crystal structure is distorted from ideal triangular symmetry, the magnetic structure is that expected for an undistorted triangular lattice antiferromagnet. The  magnetic ordering wavevector is close to the commensurate value ${\bf k}=(0,1/3,0)$ and the spins form a helical structure in the $ac$ plane with $\sim120^\circ$ between nearest neighbors. By simulating the magnetic structure as a function of the four independent nearest neighbor exchange interactions we show how the specific type of distortion in $\alpha$-CaCr$_2$O$_4$ is able to give rise to the observed highly symmetric magnetic order.

\section{Experimental details}

Both polycrystalline and single-crystal samples of $\alpha$-CaCr$_2$O$_4$ were synthesized. The polycrystalline samples were produced by a solid state reaction. A 1:1 molar ratio of CaCO$_3$ and Cr$_2$O$_3$ was mixed thoroughly, pressed into a pellet and annealed in N$_2$ atmosphere at $1100^\circ$C for 12 hours. The powder was then quenched in liquid nitrogen to prevent the formation of impurity phases. The single-crystal growth was performed using a high-temperature optical floating zone furnace at the Crystal Laboratory, Helmholtz Zentrum Berlin  f\"{u}r Materialien und Energie (HZB), Berlin, Germany, details of the technique will be reported elsewhere. \cite{nazmul3000} The single-crystal used for neutron scattering experiments is plate-like with shiny flat surfaces perpendicular to the $a$-axis, it has a weight of $340$ mg and a size of $3\times7\times8$ mm$^3$. This crystal consists of three crystallographic twins rotated by $60^{\circ}$ with respect to each other about the shared $a$-axis. Smaller single-crystal pieces were used for the bulk properties measurements.

Bulk properties measurements were performed at the Laboratory for Magnetic Measurements, HZB. Heat capacity was measured on a small single-crystal sample of 15.6 mg using a Physical Properties Measurement System (PPMS), Quantum Design over the temperature range of $2.1 \leq T \leq 300$ K. High temperature DC magnetic susceptibility ($300 \leq T \leq 1000$ K) was measured in a 1 T field on an unoriented crystal using the PPMS with vibrating sample magnetometer and oven options. Low temperature ($2.1 \leq T \leq 300$ K) susceptibility at 1 T  and magnetisation up to 5 T were measured both parallel and perpendicular to the $a$-axis ($\chi_{\text{a}||}$ and $\chi_{\text{a}\perp}$) using a superconducting quantum interference device (Magnetic Property Measurement System, Quantum Design). The single-crystal had a weight of 1.81 mg and was fixed with GE varnish to a plastic sample stick. Both field cooled and zero field cooled susceptibility were measured but no significant difference was observed.

X-ray powder diffraction was measured on the ID31 high-resolution powder-diffraction beamline at the European Synchrotron Radiation Facility (ESRF), Grenoble, France. A wavelength of $\lambda=0.39983$ \AA\ ($E=31.18$ keV) was used, and data were collected at a number of temperatures in the range $10 \leq T\leq 260$ K. Neutron powder patterns were measured on two instruments. First on the D20 high-intensity two-axis diffractometer at the Institute Laue Langevin (ILL), Grenoble, France. A Ge monochromator was used to select a wavelength of $\lambda=1.869$ \AA, and data were collected over the temperature range of $10 \leq T\leq 42.6$ K in 1 K steps. Measurements were also performed on the E9 fine-resolution powder diffractometer at the BER II reactor, HZB. A Ge monochromator selected a wavelength of $\lambda=1.799$ \AA\ and data were collected at 2.1 K and 50 K. 

Single-crystal neutron diffraction was performed on the large (340 mg) crystal using the E5 four-circle single-crystal diffractometer at the BER II reactor, HZB. A pyrolytic graphite (PG) monochromator selected an incident wavelength of $\lambda=2.36$ \AA. The instrument is equipped with a two-dimensional position-sensitive $^3$He detector consisting of 36 x 36 pixels and for each Bragg reflection a 36-step scan of the crystal angle ($\omega$) was performed. A total of 616 magnetic and 423 nuclear Bragg reflections were collected at 6 K for a total counting time of five minutes each.

Spherical neutron polarimetry was performed on the same single-crystal using the TASP triple axis spectrometer at the Paul Scherrer Institute (PSI), Switzerland. The wavelength $\lambda=3.2$ \AA\ was selected by a PG monochromator and benders were used to polarize and analyse the neutron beam along the vertical $z$-axis. 
All measurements took place at a sample temperature of 1.5 K. The  MuPAD option was installed to control the polarisation direction of the incoming neutron beam (${\bf P}_\text{i}$) and also to analyse the scattered beam along any final polarisation direction (${\bf P}_\text{f}$). The beam polarisation was measured to be 0.883(10) for all directions of neutron polarization. The conventional right-handed Cartesian coordinate system was used to describe the polarization with the $x$ axis parallel to the wavevector transfer $\bf Q$, $y$ in the horizontal scattering plane perpendicular to $\bf Q$, and $z$ vertical. There are a total of 36 cross sections that can be measured for every Bragg reflection, since both ${\bf P}_\text{i}$ and ${\bf P}_\text{f}$ can be along $\pm x$, $\pm y$ and $\pm z$. For most Bragg peaks only 18 cross sections were measured by setting the incoming beam polarisation to the positive value, while for a few peaks all 36 cross sections were measured to check chirality. Since the measured intensities were rather small, 
the background was also measured for all cross sections close to each Bragg peak. The polarisation matrices were calculated from the background corrected cross sections:  $\mathcal{P}_\text{ij}=(\sigma^{+i\rightarrow+j}-\sigma^{+i\rightarrow-j})/(\sigma^{+i\rightarrow+j}+\sigma^{+i\rightarrow-j})$, this method has the advantage of canceling out positioning errors and any wavevector dependent attenuation since only the ratio of the intensities matters. A total of 23 reflections were measured with the crystal oriented in three different scattering planes $(h,k,0)$, $(h,k,3k/2)$ and $(h,k,-3k/2)$. 
All cross sections were counted for three minutes each.

\section{Results}
\subsection{Bulk properties measurements}
\begin{figure}[!htb]
  \centering
  \leavevmode
  \includegraphics[width= 86 mm]{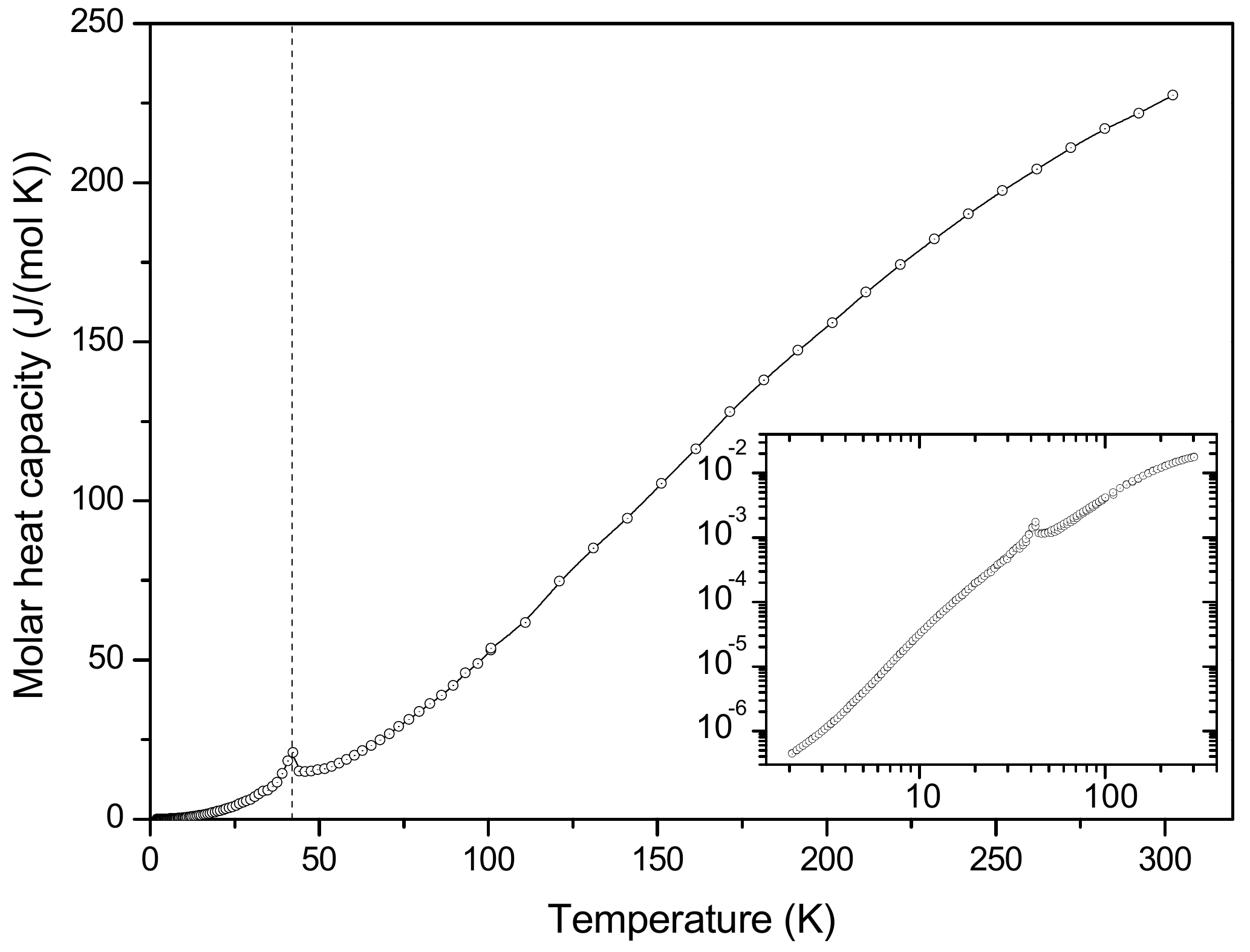}
  \caption{Heat capacity showing the magnetic phase transition at $T_\text{N}=42(1)$ K. The inset shows the same data plotted on a log-log scale.}
  \label{fig:cp}
\end{figure}

The specific heat of $\alpha$-CaCr$_2$O$_4$ as a function of temperature is shown in Fig.\ \ref{fig:cp}. A sharp peak is observed at $T_\text{N}=42(1)$ K  in agreement with Ref.\ \onlinecite{Chapon2011}. It is attributed to the onset of long-range antiferromagnetic order. No other sharp features were observed in this temperature range indicating the absence of further magnetic or structural transitions. 

\begin{figure}[!htb]
  \centering
  \leavevmode
  \includegraphics[width= 86 mm]{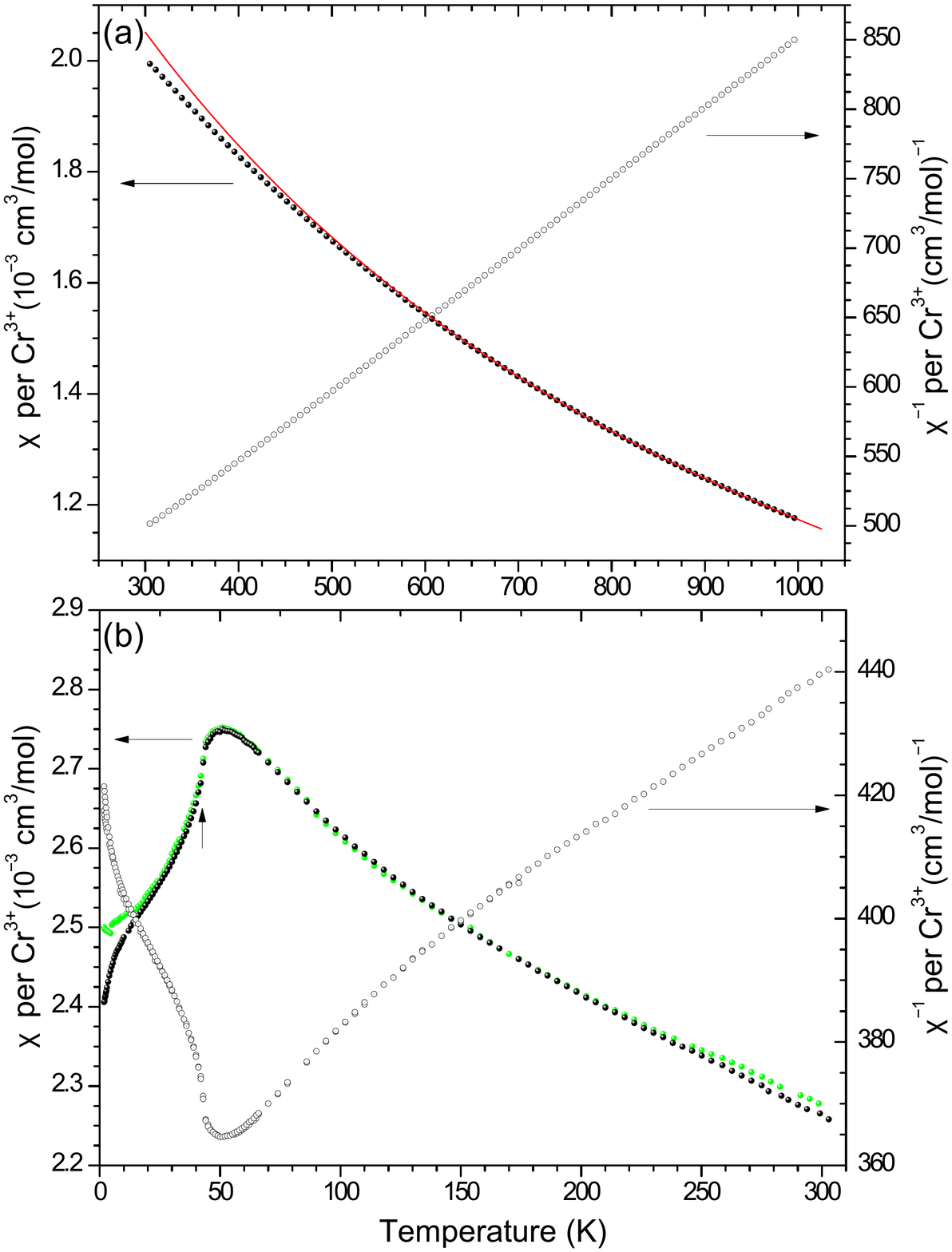}  
  \caption{DC magnetic susceptibility per Cr$^{3+}$ ion measured on a single-crystal sample in a 1 T magnetic field.
    (a) left axis: high temperature susceptibility of unoriented sample (filled data points), right axis: inverse susceptibility (open data points), the red line is the Curie--Weiss fit. 
    (b) Low temperature magnetic susceptibility; green data points, $\chi_{\text{a}||}$; black points, $\chi_{\text{a}\perp}$. 
    The vertical arrow shows the inflection point at $T_N$. Right axis: inverse susceptibility ($\chi^{-1}_{\text{a}\perp}$) displayed by the open data points.}
  \label{fig:suscSC}
\end{figure}

Figure \ref{fig:suscSC}(a) shows the high temperature susceptibility ($300 \leq T\leq 1000$ K), which reveals Curie--Weiss behavior. The data in the temperature range 800 K to 1000 K were fitted to the Curie--Weiss law:
\begin{align}
  \chi_\text{CW}=\frac{C}{T-\Theta_\text{CW}}+\chi_0,
\end{align}
where $\Theta_\text{CW}$ is the Curie--Weiss temperature and $C$ is the Curie constant measured in units of cm$^3$mol$^{-1}$K$^{-1}$. $\chi_0$ is a temperature independent background, which accounts for the paramagnetic Van Vleck susceptibility and diamagnetic core susceptibility as well the background from sample stick and Al-cement (which is less than 0.5\% of the measured susceptibility at 1000 K). 
The fitted value of the Curie-Weiss temperature is $\Theta_\text{CW}=-564(4)$ K and the effective moment which is given by $\sqrt{8 C} \mu_\text{B}$ was found to be $\mu_\text{eff}=3.68(1) \mu_\text{B}$. This is close to the expected spin only value $3.87 \mu_\text{B}$, assuming that the orbital angular momentum is completely quenched. The average value of the exchange interactions in the triangular layers can be estimated from the Curie-Weiss temperature using the mean-field approximation and is given by $k_B\Theta=S(S+1) z J_\text{mean}/3$, where $z$ is the number of nearest neighbours (6 for the triangular lattice) and the interlayer interactions have been neglected. The extracted average intralayer coupling in $\alpha$-CaCr$_2$O$_4$ is antiferromagnetic with strength $J_\text{mean}=6.48(5)$ meV.

The susceptibility at low temperatures ($2.1 \leq T\leq 300$ K) for applied field parallel and perpendicular to $a$ is shown in Fig.\ \ref{fig:suscSC}(b). The data deviate from the Curie--Weiss law and show a broad maximum at 50 K for both crystal orientations. This feature can be explained by the onset of low dimensional antiferromagnetic correlations in the triangular plane. The susceptibility decreases suddenly at the same temperature where heat capacity shows a phase transition.  The large drop reveals the onset of long-range antiferromagnetic order and the best estimate for the transition temperature is $T_\text{N}=43$ K. 

\subsection{Crystal structure}
\begin{figure}[!htb]
  \centering
  \leavevmode
  \includegraphics[width= 86 mm]{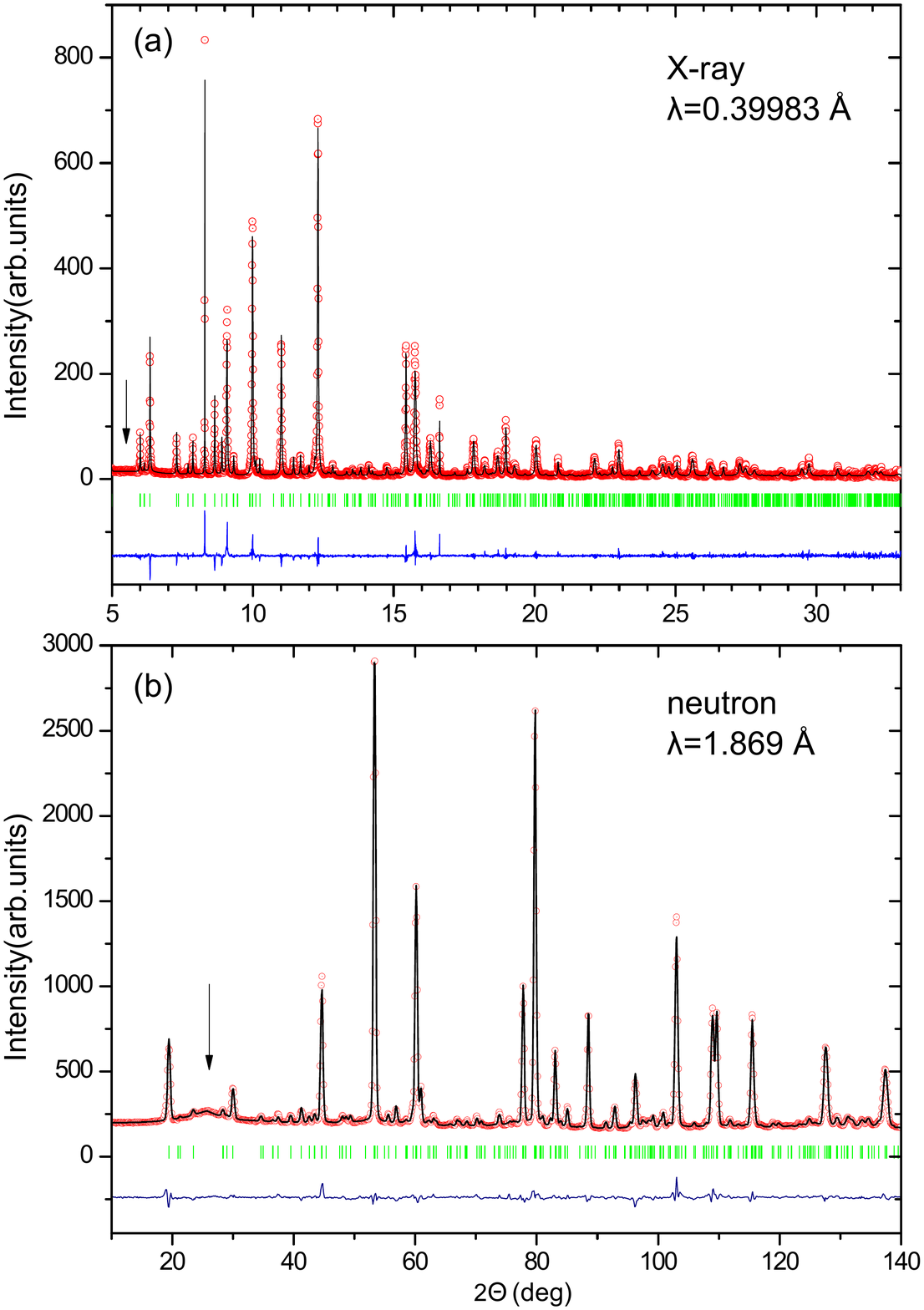}  
  \caption{Powder diffraction patterns along with Rietveld refinements. The open red symbols are the data, the black curve is the fit, the green bars indicate the nuclear Bragg peak positions and the lower blue line gives the difference between fit and data.
    (a) X-ray powder diffraction measured on ID31 at 50 K, the strongest peak (2,0,0) at $3^\circ$ is excluded from refinement.
    (b) Neutron powder diffraction measured on D20, ILL at 50 K, the vertical arrow indicates the broad bump which is of magnetic origin and therefore missing on the X-ray pattern.} 
  \label{fig:nucrefine}
\end{figure}

The crystal structure of $\alpha$-CaCr$_2$O$_4$ was refined from powder neutron diffraction data (measured on E9 and D20) along with X-ray powder diffraction data (measured on ID31) collected above the antiferromagnetic phase transition at 50 K (Fig.\ \ref{fig:nucrefine}). The structural refinement was performed using the FullProf software. \cite{Rodriguez-Carvajal1993} All the peaks could be indexed in the space group $Pmmn$ in agreement with previous powder measurements \cite{Pausch1974,Chapon2011} and no detectable impurity was observed. 
There is a broad bump around $2\theta=25^\circ$ in the neutron diffraction pattern that is absent from the X-ray pattern which, due to the insensitivity of X-rays to magnetic moments, must be of magnetic origin. This provides further evidence for two-dimensional magnetic correlations above $T_N$ at 50 K in agreement with the temperature of the maximum in the susceptibility (Fig.\ \ref{fig:suscSC}(b)). The high resolution synchrotron data show that Bragg peaks with $(0,k,l)$ are broadened (this was not observable in the neutron data because of the poorer resolution). This could be due to small planar rotations between adjacent layers perpendicular to the $a$ axis.

Table \ref{tab:rlu} lists the fitted atomic positions from the different refinements. Our values are in agreement with those from the previous room temperature measurement of Ref.\ \onlinecite{Pausch1974}. The quality of fit is given by the $R_F$-factor values: $R_F=\sum|F_\text{OBS}-F_\text{CALC}|/\sum |F_\text{OBS}|$, where $F_\text{OBS}$ and $F_\text{CALC}$ are the observed and calculated structure factors. They are: 0.035 for ID31 at 50 K, 0.037 for D20 at 47 K, 0.035 for E9 at 50 K and 0.031 fr E9 at 2.1 K. The refined model confirms that $\alpha$-CaCr$_2$O$_4$ consists of distorted triangular layers with two symmetry inequivalent Cr$^{3+}$ ions in the $bc$ plane. There are a total of four intraplanar nearest neighbor distances varying between 2.882 \AA\ and 2.932 \AA\ at 2.1 K, which correspond to four inequivalent nearest neighbour direct exchange paths (see Fig.\ \ref{fig:dist}(b)). The interactions follow a complex pattern, where equivalent exchange paths form two different zig-zags and two different chains in the triangular plane. Cr--O--Cr angles vary between $91.0^\circ$ and $95.7^\circ$ confirming that the superexchange interactions via oxygen are weak while the large interplane distances which are 5.529 \AA\ (Cr(1)--Cr(1)) and 5.379 \AA\ (Cr(2)--Cr(2)) suggest weak interplane interactions.
\begin{ruledtabular}
\begin{table}[!htb]
	\centering
	\caption{Refined crystallographic parameters using space group $Pmmn$. The lattice parameters were determined from the high resolution ID31 X-ray powder data at 50K. The atomic position listed are; first row, ID31 powder data at 50 K ($R_F=0.035$); second row, refinement of D20 neutron powder data at 47 K ($R_F=0.037$); third row, refinement of E9 powder data at 50 K ($R_F=0.035$); fourth row, E9 powder data at 2.1 K ($R_F=0.031$).}
	\label{tab:rlu}
	\begin{tabular}{p{0.4in}p{0.4in}p{0.8in}p{0.8in}p{0.8in}}    
    \multicolumn{3}{l}{Unit cell dimensions:}&\multicolumn{2}{l}{$a= 11.0579(2)$ \AA}\\
    \multicolumn{3}{l}{                     }&\multicolumn{2}{l}{$b= \;\;\,5.8239(2)$ \AA}\\
    \multicolumn{3}{l}{                     }&\multicolumn{2}{l}{$c= \;\;\,5.0553(1)$ \AA}\\
    \\
    Atom\B  &	Site&	$x$	     &	$y$	        &	$z$				\\        
    \hline
    Ca(1)\T & 2$a$&	3/4	      &	3/4	        &	0.3512(4)	\\
    	      &		  &		        &		          & 0.3494(10)\\    	    	
    	      &		  &		        &		          & 0.3521(12)\\    	    	    	
    	      &		  &		        &		          & 0.3532(12)\\    	    	    	    	
    \hline	
    Ca(2)	  &	2$b$&	1/4	      &	3/4	        &	0.0385(4)	\\
    	      &			&			      &			        &	0.0349(10)\\    	    	
    	      &			&			      &			        &	0.0384(11)\\    	    	    	
    	      &			&			      &			        &	0.0410(11)\\    	    	    	    	
    \hline	                                   
    Cr(1)	  &	4$d$&	1/2	      &	1/2	        &	1/2	      \\
    \hline
    Cr(2)	  &	4$f$&	0.4932(1)	&	1/4	        &	0.0046(4)	\\
    	      &			&	0.4921(4)	&			        &	0.0050(9) \\    	    	
    	      &			&	0.4917(5)	&			        &	0.0059(11)\\    	    	    	
    	      &			&	0.4931(6) &			        &	0.0057(12)\\    	    	    	   	
    \hline	
    O(1)	  &	4$f$&	0.4022(5)	&	1/4	        &	0.3365(14)\\    	
    	      &			&	0.4022(4)	&			        &	0.3394(16)\\    	    	
    	      &			&	0.4023(4)	&			        &	0.3397(8) \\    	    	    	
    	      &			&	0.4024(4)	&			        &	0.3410(8) \\    	    	    	    	
    \hline	
    O(2)	  &	4$f$&	0.5904(5)	&	1/4	        &	0.6825(14)\\    	
    	      &			&	0.5891(3)	&			        &	0.6771(6)	\\    	    	
    	      &			&	0.5901(4)	&			        &	0.6758(8) \\    	    	    	
    	      &			&	0.5906(4)	&			        &	0.6762(8) \\    	    	    	    	
    \hline	                                   
    O(3)	  &	8$g$&	0.5989(3)	&	0.4996(8) 	&	0.1665(12)\\    	
    	      &			&	0.6007(2)	&	0.4963(5)		&	0.1653(5)	\\    	    	
    	      &			&	0.6003(2)	&	0.4955(6)		&	0.1647(7) \\    	    	        	
    	      &			&	0.6003(2)	&	0.4966(6) 	&	0.1657(7) \\    	    	  
	\end{tabular}
\end{table}
\end{ruledtabular}

\begin{figure}[!htb]
  \centering
  \includegraphics[width= 86 mm]{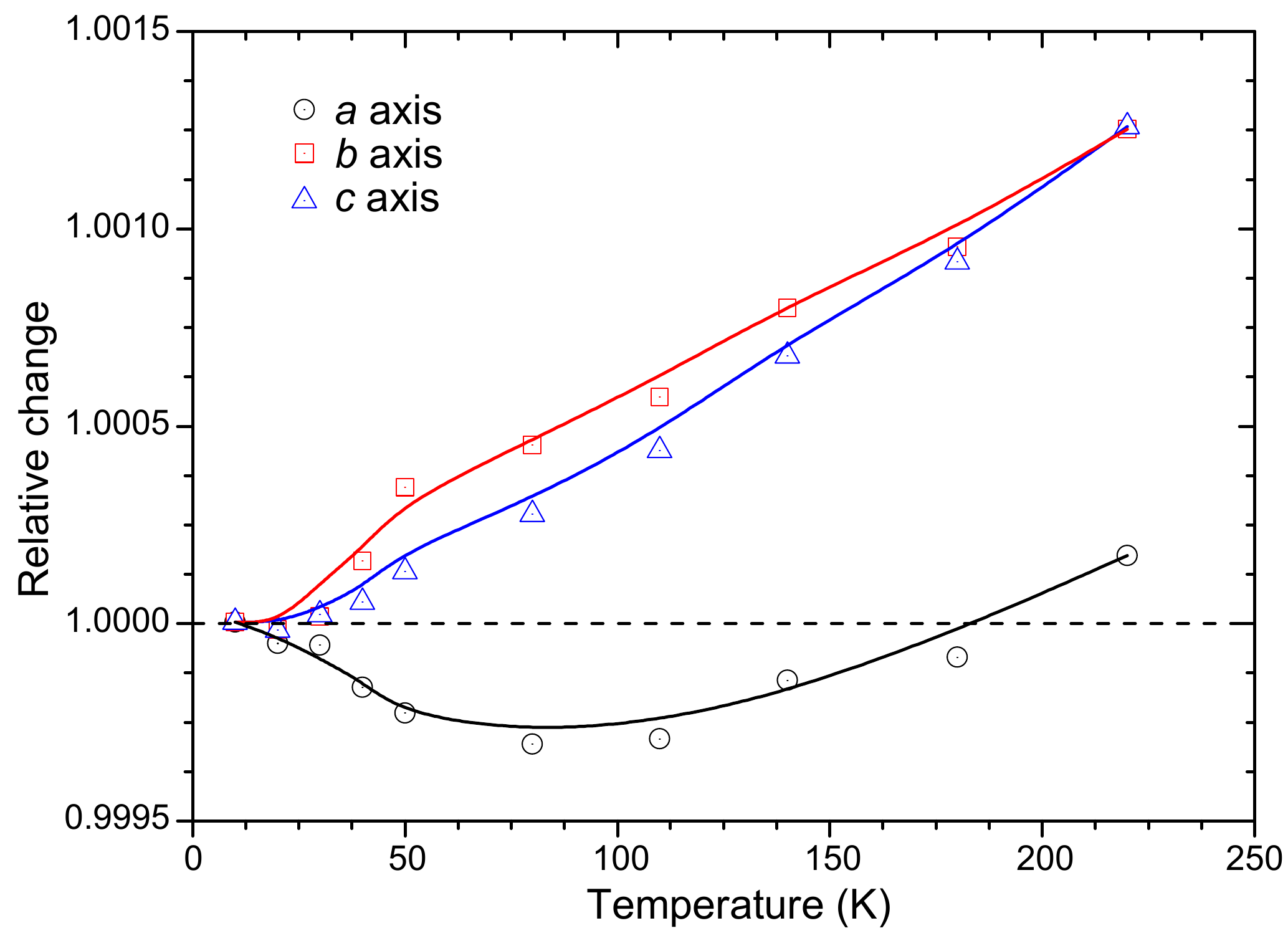}  
  \caption{Temperature dependence of the lattice parameters obtained from ID31 data. The lines are guides to the eye.}
  \label{fig:seqlattice}
\end{figure}

The temperature dependence of the crystallographic parameters were determined by sequential refinement of the data collected on ID31 over a temperature range between 10 K and 300 K. No significant change in the relative atomic positions was observed, and in particular there is no sudden change in the structure at the N\'{e}el temperature. For comparison see the atomic positions at 2.1 K and 50 K determined from E9 refinement in Table \ref{tab:rlu}. Furthermore the lattice parameters were found to vary smoothly as shown in Fig.\ \ref{fig:seqlattice}. The $b$ and $c$ lattice constants increase gradually with increasing temperature, while the $a$ lattice constant is smallest at 70 K and shows negative thermal expansion at lower temperatures. There is no indication of any symmetry   change over this temperature range, even a subtle distortion can be ruled out since the linewidth remains constant between 10 K and room temperature. This is in contrast to CuCrO$_2$, \cite{Kimura2009} which unlike $\alpha$-CaCr$_2$O$_4$ has perfect triangular symmetry at high temperatures and undergoes a structural distortion at the N\'{e}el temperature.

Single-crystal diffraction on the E5 instrument was used to collect the nuclear Bragg peak intensity at 6 K.
The Racer program was used to integrate the reflections, as described in Ref.\ \onlinecite{Wilkinson1988}. Since our single-crystal has a large mosaic spread ($\sim2^\circ$) with several small grains at slightly different orientations, in a few cases it was not possible to completely exclude the intensity of these grains, which unfortunately impacted on the data quality.

\begin{figure}[!htb]
\centering
  \leavevmode
  \includegraphics[width= 86 mm]{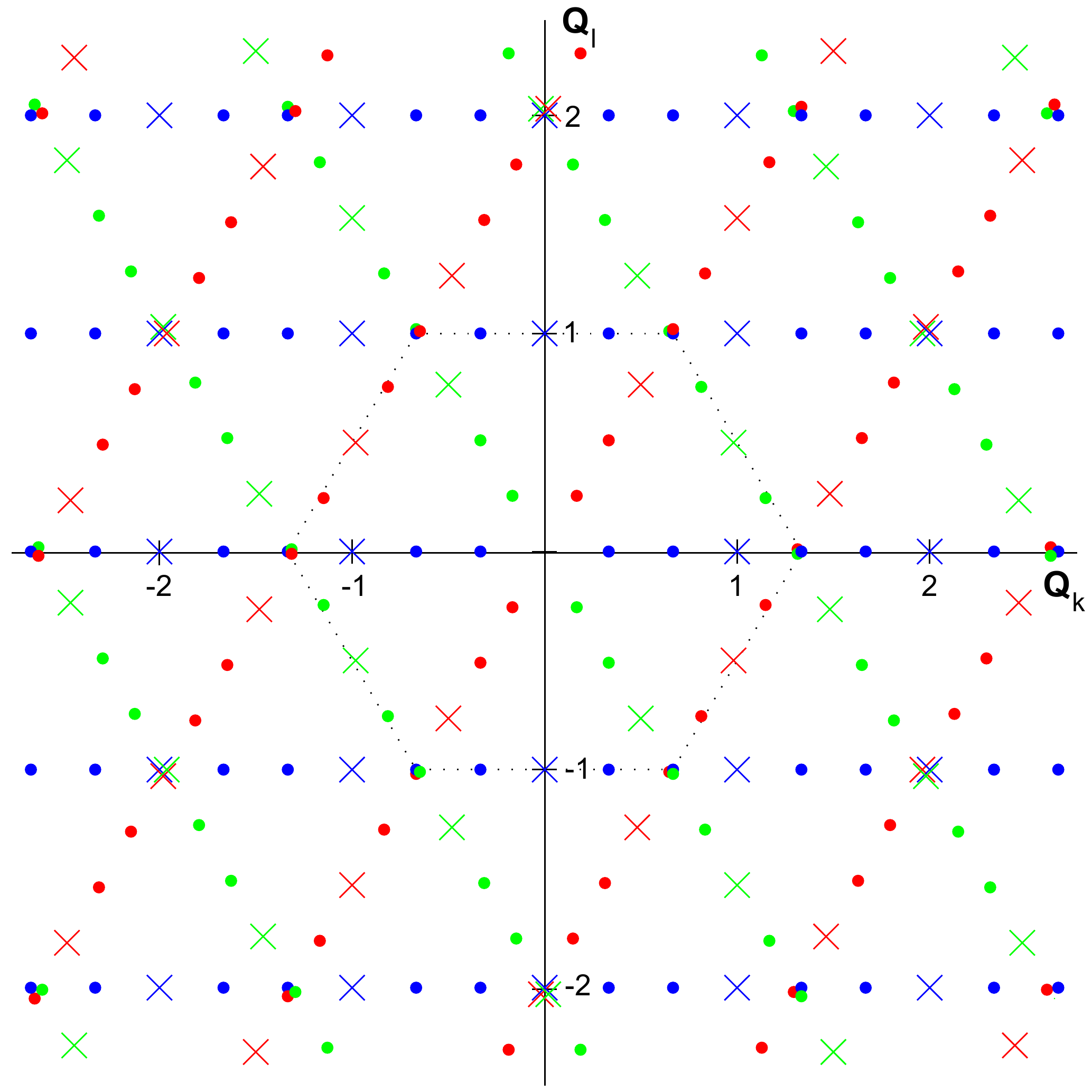}
  \caption{The $(1,k,l)$ reciprocal space plane where distortion is exaggerated for better visibility. The crosses give the position of the nuclear reflections while the dots show all possible magnetic reflections. Each color correspond to a different twin. The dashed line indicates the border of magnetic Brillouin zone.}
  \label{fig:kl}
\end{figure}

Another problem that arises with the single-crystal data is twinning. Since the three-fold symmetry of the triangular planes is weakly distorted, three structural twins can exist rotated with respect to each other by $60^\circ$ in the $bc$ plane, while sharing the out-of-plane $a$-axis. Figure \ref{fig:kl} shows all the possible nuclear reflections for all three twins in the $(1,k,l)$ plane. Some reflections arise from one twin only e.g. $(1,1,0)$; and can be used to determine the ratio of the twins by simply taking the intensity ratio of equivalent reflections of the three twins. Other reflections have overlapping contributions from all three twins which cannot be resolved on E5 e.g. $(1,0,2)$ which is labeled $(1,2,1)$ and $(1,2,-1)$ in the other two twins. 

A single intensity list was generated containing reflections of all three twins and each reflection was labeled by the twin(s) to which it belonged. One index was assigned to the untwinned peaks while three indices were assigned to the twinned reflections (each in the coordinate system of that twin). 
Using the atomic coordinates obtained from the E9 diffraction data at 2.1 K, this dataset was refined to obtain the scale factor and the volume ratios of the twins. 
The three twins were found to have non-equal weight (see Table \ref{tab:twin}), and their extracted ratios were essential to the magnetic structure refinement of the low temperature single crystal data which is described in the next section.
\begin{ruledtabular}
\begin{table}[!htb]
	\centering

	\caption{Twin ratio of the single-crystal sample, determined from single-crystal diffraction on E5 at 6 K.}
	\label{tab:twin}

	\begin{tabular}{p{2.2cm}p{2.2cm}p{2.2cm}}
		Twin 1&Twin 2 (+60$^\circ$)&Twin 3 (+120$^\circ$)\\
		\hline
		 64(2)\% & 19(2)\% & 17(3)\% \\
	\end{tabular}
\end{table}
\end{ruledtabular}
\subsection{Magnetic structure}

Several different neutron diffraction techniques were necessary to determine unambiguously the magnetic structure of $\alpha$-CaCr$_2$O$_4$; these were powder diffraction, single-crystal diffraction and spherical neutron polarimetry. We will discuss the results of each and show how together they provide a complete picture of the magnetic order. 

\begin{figure}[!htb]
  \centering
  \includegraphics[width= 86 mm]{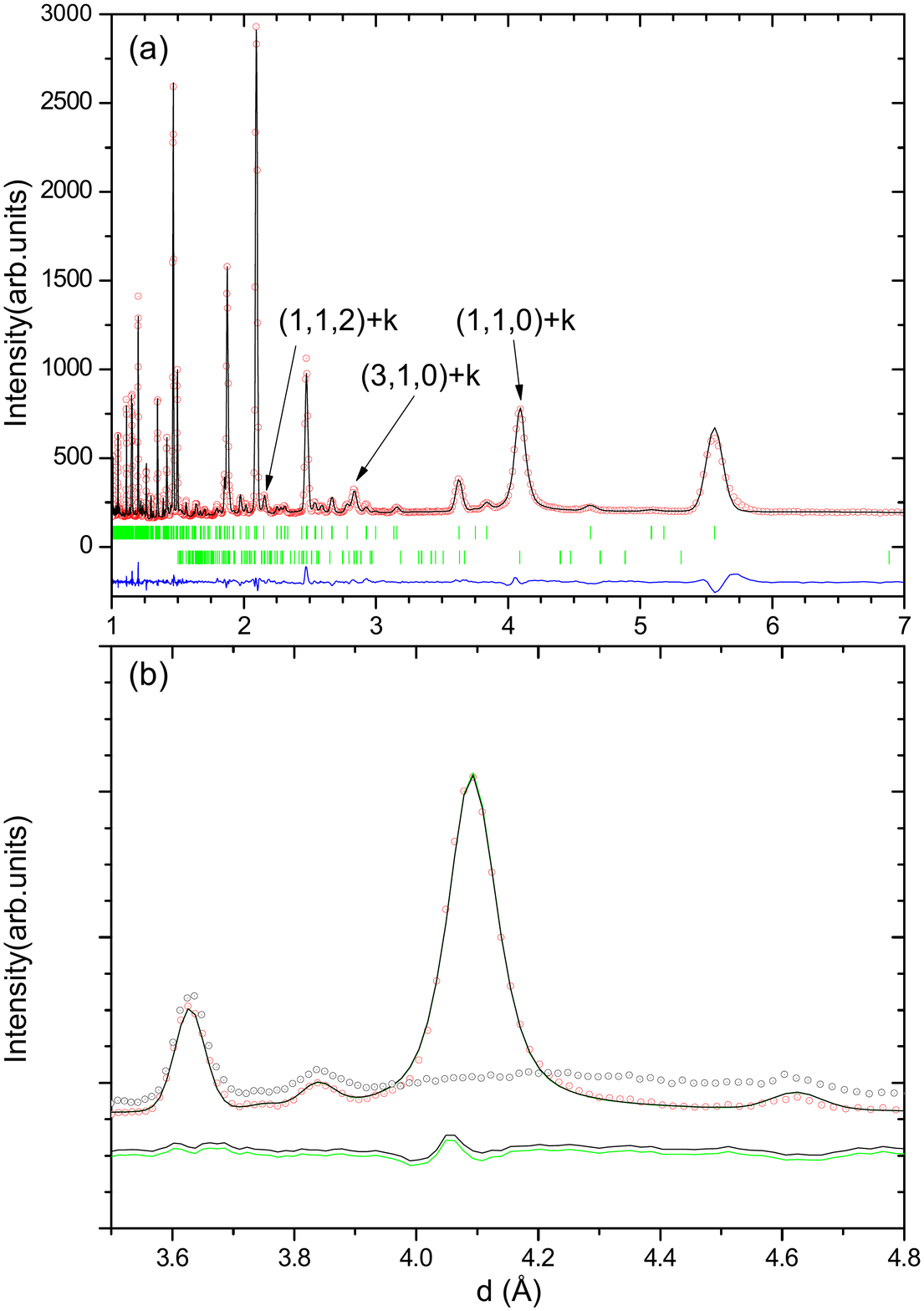}  
  \caption{Powder diffraction data measured on D20 at 10 K are displayed along with the Rietveld refinement assuming helical magnetic order in the $ac$ plane. 
    (a) The open red symbols are the data, the black curve is the fit, the upper and lower green bars indicate the nuclear and magnetic Bragg peak positions respectively, and the lower blue line gives the difference between fit and data. 
    (b) Detail of the $(1,1,0)+{\bf k}$ magnetic peak along with the 50 K data (open black symbols). The green and blue solid lines are the difference between the data and the fit with $\bf k$ fixed to (0,1/3,0) and refined to (0,0.332(3),0) respectively.} 
  \label{fig:magref}
\end{figure}

The neutron powder diffraction pattern from D20 measured well below $T_N$ at 10 K is shown in Fig.\ \ref{fig:magref}. Several new peaks are observed which are attributed to the magnetic order, the strongest is at $d=4.093$ \AA\ where the broad bump was found in the 50 K diffraction pattern. 
The new peaks were indexed using the k-search program of the FullProf suite, the magnetic ordering wavevector $\bf k$ was found to be close to the commensurate value ${\bf k}=(0,1/3,0)$ in agreement with Ref.\ \onlinecite{Chapon2011}, see Table \ref{tab:ksearch}.  Only magnetic reflections where $h$ is odd are observed, which implies antiferromagnetic ordering between successive planes along $a$ (since the unit cell contains two Cr$^{3+}$ layers stacked along $a$). All possible magnetic reflections in the $(1,k,l)$ plane generated by ${\bf k}=(0,1/3,0)$ are shown in Fig.\ \ref{fig:kl}. However, only magnetic peaks where the three twins overlap have magnetic intensity e.g.\ $(1,4/3,0)$; this makes the magnetic refinement challenging.
\begin{ruledtabular}
\begin{table}[!htb]
	\centering
	
	\caption{The magnetic peaks indexed assuming ${\bf k}=(0,1/3,0)$ are listed along with their observed and calculated $d$-spacings.}
	\label{tab:ksearch}
	
	\begin{tabular}{p{2.2cm}p{2.2cm}p{2.2cm}}
		$d_\text{obs}$ (\AA)& index     & $d_\text{cal}$ (\AA)\\
		\hline
		4.093     & $(1,1,0)+{\bf k}$ & 4.084      \\
		2.833     & $(3,1,0)+{\bf k}$ & 2.832      \\
		2.154     & $(1,3,0)-{\bf k}$ & 2.154      \\
		1.642     & $(1,3,3)-{\bf k}$ & 1.645      \\
	\end{tabular}
\end{table}
\end{ruledtabular}

The magnetic structures allowed for $\alpha$-CaCr$_2$O$_4$ can be derived from symmetry analysis, see Ref. \onlinecite{Chapon2011}. 
A number of magnetic structures are possible including helical, cycloidal, ellipsoidal and sinusoidal with the spin moments pointing along different directions or rotating in different planes. 
\begin{ruledtabular}
\begin{table*}[!htb]
	\centering

	\caption{Refinement of the magnetic structure for spins pointing in various directions or planes for the two data sets (D20 and E5). The column headings define the axes along which the magnetic moment is nonzero and $m_x$, $m_y$ and $m_z$ are the moment sizes along these axes. All moment lengths are in units of $\mu_\text{B}$. The $R$-factors are also included in order to compare the quality of the fits, $R_\text{B}$ is determined from the integrated intensities: $R_B=\sum|I_\text{OBS}-I_\text{CALC}|/\sum |I_\text{OBS}|$. Helical and sinusoidal models give the same R-factors.}
	\label{tab:maghel}
	
	\begin{tabular}{p{1.3in}p{0.5in}p{0.5in}p{0.5in}|p{0.5in}p{0.5in}|p{0.5in}p{0.5in}|p{0.5in}p{0.5in}p{0.5in}}
                        &              &\multicolumn{2}{c}{$ab$}&\multicolumn{2}{c}{$ac$}&\multicolumn{2}{c}{$bc$}&$a$&$b$&$c$\\
		\hline
	  D20 data ($T=10$ K):& $m_\text{x}$ & 2.62(5)  & 2.18(1)  & 2.50(4)  & 2.24(2)  &          &          &2.88(2)&       &       \\
	                      & $m_\text{y}$ & 1.39(11) & $=m_x$   &          &          & 2.40(8)  & 2.35(2)  &       &3.11(2)&       \\
                        & $m_\text{z}$ &          &          & 1.88(7)  & $=m_x$   & 2.23(10) & $=m_y$   &       &       &3.30(3)\\
                        & $R_\text{B}$ & 0.045    & 0.066    & 0.032    & 0.068    & 0.212    & 0.209    &0.067  &0.190  &0.255  \\
		\hline
	  E5 data ($T=6$ K):  & $m_\text{x}$ & 2.34(3)  & 2.09(1)  & 2.26(2)  & 2.12(1)  &          &          &2.99(2)&       &       \\
	                      & $m_\text{y}$ & 1.78(4)  & $m_x$    &          &          & 1.90(14) & 2.10(13) &       &3.09(3)&       \\
	                      & $m_\text{z}$ &          &          & 1.97(3)  & $m_x$    & 2.34(13) & $m_y$    &       &       &3.14(2)\\  
                        & $R_\text{F}$ & 0.338    & 0.341    & 0.321    & 0.322    & 0.392    & 0.392    &0.379  &0.400  &0.401  \\
	\end{tabular}
\end{table*}
\end{ruledtabular}
Table \ref{tab:maghel} gives the fitted magnetic moments along the crystallographic axes. The refined phase between the Cr(1) and Cr(2) sites was close to $-2\pi /3$ , which was afterwards fixed to this value for all the refinements. The moment size on the two inequivalent Cr$^{3+}$ sites were refined independently, but found to have the same size within standard deviation. Therefore they were constrained to be equal for all subsequent refinements. As can be observed, the results of the refinement are ambiguous since several structures have similar $R$-values. The powder data favors models where the spins point either in the $ab$ or $ac$ planes; however in these cases it is not possible to distinguish between helical or sinusoidal order because the phase difference between the two prefactors cannot be refined. Nevertheless the sinusoidal structures can probably be excluded indirectly as shown in Ref.\ \onlinecite{Chapon2011}. For example the refined moment lengths for the $ac$ structure are $m_x=2.50(4)\mu_\text{B}$ and $m_z=1.88(7)\mu_\text{B}$. Assuming sinusoidally modulated order, the maximum moment length would exceed the spin only magnetic moment of Cr$^{3+}$: $\sqrt{m_x^2+m_z^2}>g S$, orbital momentum is quenched for Cr$^{3+}$ in an octahedral environment and $g$ is the Land\'e $g$-factor ($g\approx 2$ for spin-only magnetic moment). On the other hand the maximum size of the moment in the helical structure would be the $x$ component $2.50 \mu_B$.

The refined value of the $\bf k$-vector was $(0,0.332(3),0)$ from D20 at 10 K and $(0,333(5),0)$ from E9 at 2.1 K. According to the refinement the ${\bf k}$-vector is equal to $(0,1/3,0)$ within standard deviation, the difference between the diffraction patterns generated by these two $\bf k$ values is negligible (see Fig.\ \ref{fig:magref}(b)). Our results are also in agreement with the more precise incommensurate value of (0,0.3317(2),0) published previously\cite{Chapon2011}. In the case of helical structures, the phase of $-2\pi /3$ together with ${\bf k}=(0,1/3,0)$ means that the angle between all intraplanar nearest neighbor spins is $120^\circ$. This structure is expected only for the ideal triangular lattice where all nearest neighbor exchange interactions are equal and Heisenberg. It is therefore surprising to find it on the distorted triangular lattice of $\alpha$-CaCr$_2$O$_4$.

As a complementary study, we performed single-crystal diffraction at 6 K in the magnetic phase. The twin ratio and scale factor were fixed to the values obtained from the single-crystal nuclear peaks and the data were refined with the same set of possible magnetic structures as for the powder refinements. The $R_\text{F}$-factors and resulting magnetic components are listed in Table \ref{tab:maghel}. The single-crystal data is not able to distinguish between the helical and sinusoidal models either. The measurable $|M(Q)|^2$, the absolute value square of the magnetic structure factor, is identical for a helical structure and a sinusoidal structure with same moment components due to the orthorhombic symmetry. In the case of $ac$ sinusoidal structure, two magnetic S-domains can exist in the sample, if the two domains have equal weight the resulting magnitude of the structure factor is the same as in the case of the helical structure. For the plane in which the moments are aligned, the best fit gives the $ac$ plane in agreement with the powder diffraction.

\begin{figure}[!htb]
\centering
  \leavevmode
  \includegraphics[width= 86 mm]{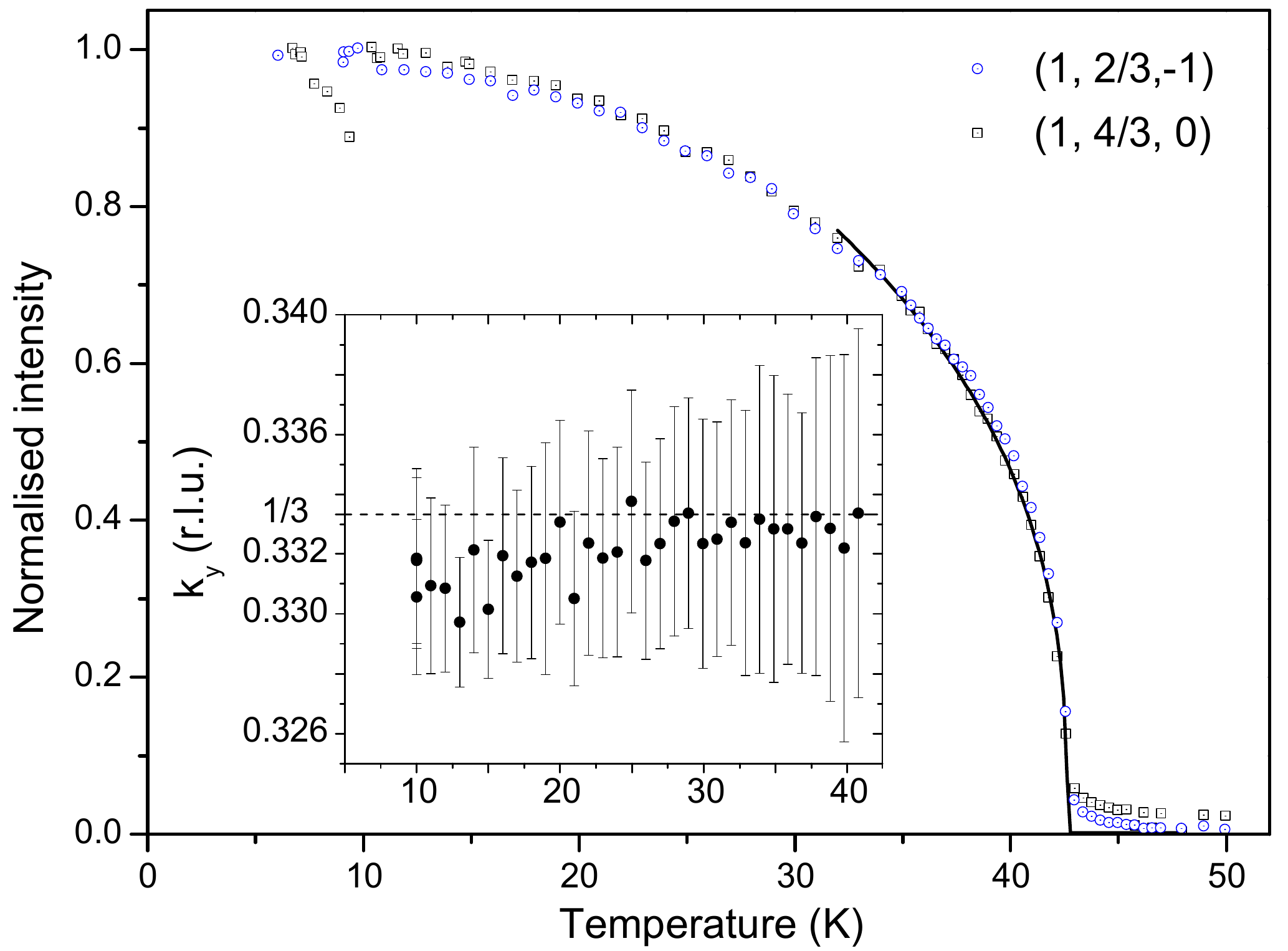}
  \caption{Temperature dependence of the (1,2/3,-1) and (1,4/3,0) magnetic reflections from the single-crystal sample, measured on E5. The black line is the fit described in the text. The inset shows the temperature dependence of $k_y$ refined from D20 data.}
  \label{fig:neel}
\end{figure}

The temperature dependence of two magnetic peaks $(1,2/3,-1)$ and $(1,4/3,0)$ were measured on the single-crystal sample in the temperature range $6 \leq T\leq 50$ K, see Fig.\ \ref{fig:neel}. The intensity is proportional to the square of the magnetisation which is the magnetic order parameter and can be fitted with $I(T)=C\cdot((T-T_\text{N})/T_\text{N})^{2\beta}$, where $C$ is an overall constant and $\beta$ is the critical exponent. The data were fitted between 32 K and 48 K and gave a transition temperature of $T_\text{N}=42.6(1)$ in agreement with the heat capacity data. The exponent is $\beta=0.1815(9)$ for both peaks. Although the function fits well, according to theory the magnetic phase transition of a stacked triangular lattice is first order, and the value of $\beta$ is not universal. \cite{Ngo2008} Several noncollinear easy axis antiferromagnets have a sinusoidal phase just below $T_\text{N}$ which precedes the helical phase, e.g.\ CuCrO$_2$ \cite{Kimura2008} and $\beta$-CaCr$_2$O$_4$. \cite{Damay2010} There is no sign of such a phase however in $\alpha$-CaCr$_2$O$_4$, the temperature dependence of the two measured Bragg peaks is smooth and the ordering wavevector remains constant up to $T_\text{N}$ (inset of Fig.\ \ref{fig:neel}).


Spherical neutron polarimetry was used to produce additional information to complement the previous neutron diffraction techniques. A number of different overlapping magnetic Bragg peaks were measured for all three twins. For example the strongest magnetic Bragg peak, $(1,4/3,0)$ indexed in the coordinate system of twin 1, overlaps with $(1,2/3,1)$ from twin 2 and $(1,-2/3,1)$ from twin 3. The $(1,4/3,0)$ peak was also measured for twins 2 and 3, where it is described as $(1,2/3,1)$ and $(1,-2/3,1)$ respectively in the coordinate system of twin 1 (see Fig.\ \ref{fig:kl}). The polarisation matrices for these three peaks labeled using the coordinate system of twin 1 are displayed in the first three rows of Table \ref{tab:polmeas}.   
\begin{ruledtabular}
\begin{table*}[!htbp]
	\centering

	\caption{Polarisation matrices measured using TASP at 1.5 K for the three strongest magnetic reflections (rows 1-3) and results of the refinement which yields a helical structure with spins lying almost entirely within the $ac$ plane (rows 4-6).} 
	\label{tab:polmeas}
	
	\begin{tabular}{r|...|...|...}
         &\multicolumn{3}{c}{(1,4/3,0)}& \multicolumn{3}{c}{(1,2/3,1) (+60$^\circ$)} & \multicolumn{3}{c}{(1,$-2/3$,1) (+120$^\circ$)} \\
                & \multicolumn{1}{c}{${\bf P}_\text{f}^x$}     & \multicolumn{1}{c}{${\bf P}_\text{f}^y$}    & \multicolumn{1}{c}{${\bf P}_\text{f}^z$}   & \multicolumn{1}{c}{${\bf P}_\text{f}^x$}   & \multicolumn{1}{c}{${\bf P}_\text{f}^y$}    & \multicolumn{1}{c}{${\bf P}_\text{f}^z$}     & \multicolumn{1}{c}{${\bf P}_\text{f}^x$}    & \multicolumn{1}{c}{${\bf P}_\text{f}^y$}    & \multicolumn{1}{c}{${\bf P}_\text{f}^z$}    \\         
        \hline                
        ${\bf P}_\text{i}^x$ &  -0.907(3)  & -0.086(9)  &  0.013(9) & -0.880(5) & -0.048(10) &   0.033(10) & -0.903(4)  & -0.036(10) &  0.004(10) \\
        ${\bf P}_\text{i}^y$ &  -0.083(9)  &  0.086(9)  & -0.037(9) &  0.054(9) &  0.372(8)  &   0.063(9)  & -0.001(9)  &  0.400(9)  & -0.110(9)  \\
        ${\bf P}_\text{i}^z$ &  -0.136(9)  & -0.029(9)  & -0.100(9) &  0.043(9) &  0.082(9)  &  -0.369(9)  & -0.014(9)  & -0.092(9)  & -0.402(9)  \\
        \hline
        ${\bf P}_\text{i}^x$ &  -0.887     &  0.000     &  0.000    & -0.887     &  0.000     &  0.000     &  -0.887    &   0.000     &  0.000    \\
        ${\bf P}_\text{i}^y$ &   0.000     &  0.098     & -0.022    &  0.000     &  0.352     &  0.081     &   0.000    &   0.407     & -0.112    \\
        ${\bf P}_\text{i}^z$ &   0.000     & -0.022     & -0.098    &  0.000     &  0.081     & -0.352     &   0.000    &  -0.112     & -0.407    \\        
  \end{tabular}
\end{table*}
\end{ruledtabular}
The observed data show strong depolarisation of the scattered beam when polarisation of the incoming beam was $y$ or $z$ due to the presence of structural twins. This makes it difficult to draw conclusions  about the axis of the spin helix by simply inspecting the data. Therefore, simulations were performed of the different models of magnetic ordering and compared to the data. The polarisation matrices were calculated using Blume-Maleev equations \cite{Chatterji2006} assuming the phase difference between Cr(1) and Cr(2) is $-2\pi /3$ and the ordering wavevector is $(0,1/3,0)$.

Polarisation matrices of three possible spin structures (helical order in the $ab$, $ac$ and $bc$ planes) were simulated for all three twins for the $(1,4/3,0)$ magnetic reflection in the coordinate system of twin 1, see Table \ref{tab:pol}. 
\begin{ruledtabular}
\begin{table*}[!htbp]
	\centering
	
	\caption{Simulations of the polarisation matrices of the $(1,4/3,0)$ reflection for helical order in the $ab$, $ac$ and $bc$ planes. The $\pm$ sign is for the two opposite vector chiral domains. The contributions of all three structural twins are listed. Columns 10--12 give the combined contribution weighted by the ratio of the twin volumes and assuming equal chiral domain ratios, the last three columns are multiplied by the finite instrumental polarisation.}
	\label{tab:pol}
	
	\begin{tabular}{lr|rrr|rrr|rrr|rrr|rrr}
                         & & \multicolumn{3}{c}{Twin 1}& \multicolumn{3}{c}{Twin 2 (+60$^\circ$)} & \multicolumn{3}{c}{Twin 3 (+120$^\circ$)} & \multicolumn{3}{c}{Together} & \multicolumn{3}{c}{$\mathcal{P}_\text{xx}$ corrected}\\
    \hline                                                                                                                                                                 
                         &         & ${\bf P}_\text{f}^x$     & ${\bf P}_\text{f}^y$ & ${\bf P}_\text{f}^z$& ${\bf P}_\text{f}^x$     & ${\bf P}_\text{f}^y$& ${\bf P}_\text{f}^z$ & ${\bf P}_\text{f}^x$      & ${\bf P}_\text{f}^y$& ${\bf P}_\text{f}^z$& ${\bf P}_\text{f}^x$      & ${\bf P}_\text{f}^y$& ${\bf P}_\text{f}^z$  &  ${\bf P}_\text{f}^x$      & ${\bf P}_\text{f}^y$& ${\bf P}_\text{f}^z$  \\
    \multirow{3}{*}{$ab$}&${\bf P}_\text{i}^x$&-1.000& 0.000& 0.000&    -1.000& 0.000 &   0.000 &      -1.000  &  0.000 &  0.000 &      -1.000  &  0.000 &  0.000   &-0.883& 0.000& 0.000\\
                         &${\bf P}_\text{i}^y$& 0.000& 1.000& 0.000&$\mp$0.977& 0.090 &   0.193 &  $\pm$0.977  &  0.090 & -0.193 &       0.000  &  0.562 &  0.005   & 0.000& 0.496& 0.005\\
                         &${\bf P}_\text{i}^z$& 0.000& 0.000&-1.000&$\mp$0.977& 0.193 &  -0.090 &  $\pm$0.977  & -0.193 & -0.090 &       0.000  &  0.005 & -0.562   & 0.000& 0.005&-0.496gl\\
    \hline                                                                                                                                                                 
    \multirow{3}{*}{$ac$} & ${\bf P}_\text{i}^x$&    -1.000& 0.000& 0.000&      -1.000 &  0.000 &  0.000  &      -1.000  &  0.000 &  0.000 &      -1.000  &  0.000 &  0.000   &-0.883& 0.000& 0.000\\
                         & ${\bf P}_\text{i}^y$ &$\mp$0.997&-0.072& 0.000&  $\pm$0.765 &  0.589 & -0.262  &  $\pm$0.765  &  0.589 &  0.262 &       0.000  &  0.105 & -0.004   & 0.000& 0.093&-0.003\\
                         & ${\bf P}_\text{i}^z$ &$\mp$0.997& 0.000& 0.072&  $\pm$0.765 & -0.262 & -0.589  &  $\pm$0.765  &  0.262 & -0.589 &       0.000  & -0.004 & -0.105   & 0.000&-0.003&-0.093\\
    \hline                                                                                                                                                                 
    \multirow{3}{*}{$bc$}&${\bf P}_\text{i}^x$&  -1.000  & 0.000 &  0.000 &      -1.000 &  0.000 &  0.000  &      -1.000  &  0.000 &  0.000 &      -1.000  &  0.000 &  0.000   &-0.883& 0.000& 0.000\\
                         &${\bf P}_\text{i}^y$&$\mp$0.647&-0.762 &  0.000 &  $\pm$0.647 & -0.762 &  0.000  &  $\mp$0.647  & -0.762 &  0.000 &       0.000  & -0.762 &  0.000   & 0.000&-0.673& 0.000\\
                         &${\bf P}_\text{i}^z$&$\mp$0.647& 0.000 &  0.762 &  $\pm$0.647 &  0.000 &  0.762  &  $\mp$0.647  &  0.000 &  0.762 &       0.000  &  0.000 &  0.762   & 0.000& 0.000& 0.673\\
  \end{tabular}
\end{table*}
\end{ruledtabular}
For the helical models the issue of chirality also needs to be addressed. In these cases the centrosymmetry of the paramagnetic space group is broken in the magnetically ordered phase, allowing the appearance of two chiral domains for each twin with opposite vector chirality. The polarisation matrix of the chiral domains differs only in the $\mathcal{P}_\text{yx}$ and $\mathcal{P}_\text{zx}$ terms, which change sign with opposite chirality, see Table \ref{tab:pol}. Because the measured values for these two elements are rather small (less than $0.136(9)$), it is assumed that the population of the two chiral domains are equal for all three twins. Fitting the weight of the chiral domains also don't improve the data significantly. In this case the measured $\mathcal{P}_\text{yx}$, $\mathcal{P}_\text{zx}$ would be zero because they cancel each other out. This, together with the presence of structural twins, explains the strong depolarisation when the incident beam has $y$ and $z$ polarisation. In order to make a comparison with the measured $(1,4/3,0)$ reflection, the contributions of the three twins weighted by their volume ratios were added together and multiplied by the beam polarisation and are listed in the last three columns of table \ref{tab:pol}. Comparison of the simulated matrices to the experimental matrix in the first three rows of Table \ref{tab:polmeas} under $(1,4/3,0)$, clearly show that the $ac$ helical structure gives the best description of the data.
\begin{figure}[!htb]
  \centering
  \includegraphics[width= 86 mm]{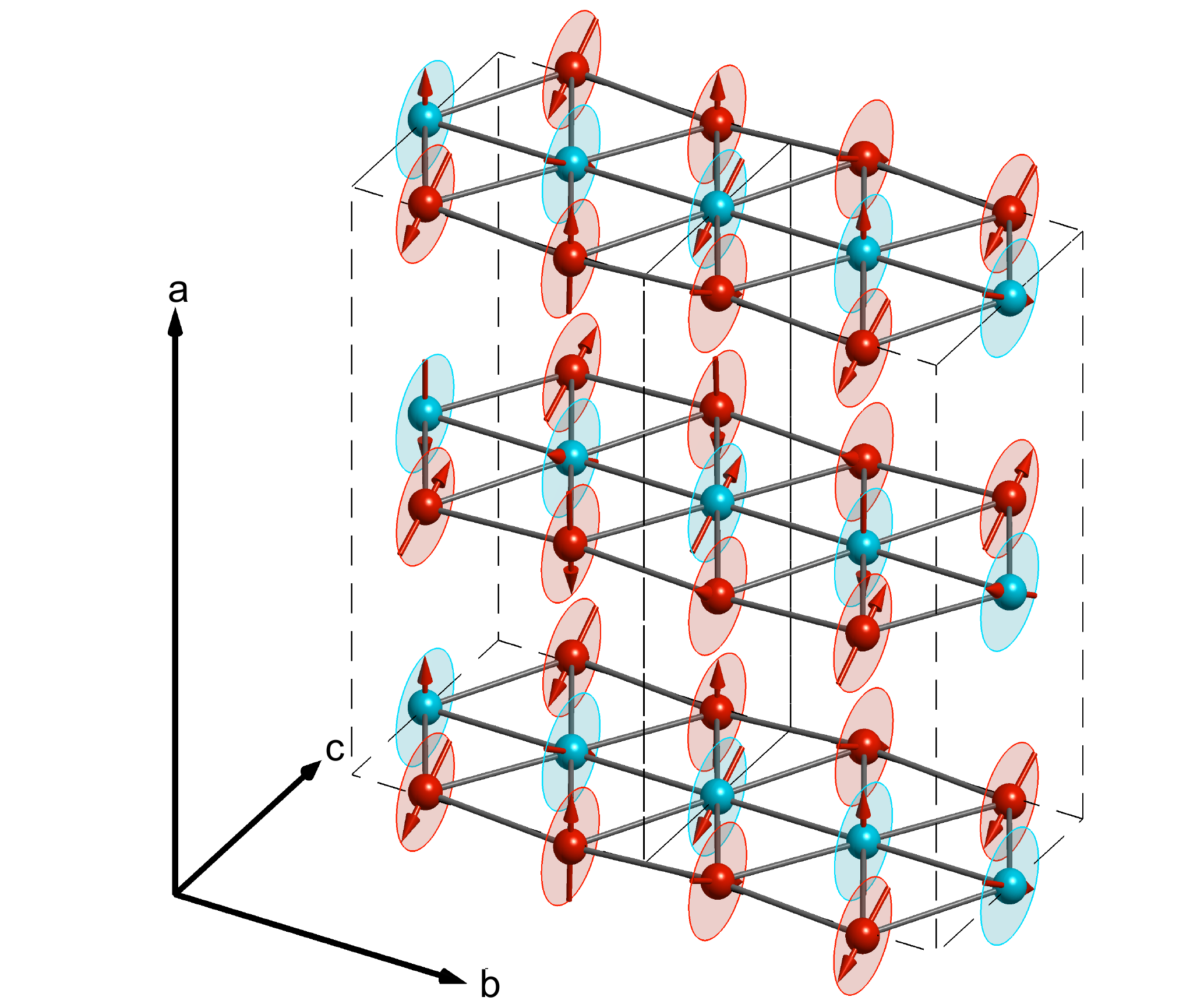}
  \caption{Magnetic structure of $\alpha$-CaCr$_2$O$_4$, determined from spherical neutron polarimetry. Cr(1) and Cr(2) are represented by the blue and red spheres respectively.}
  \label{fig:sstructure}
\end{figure}

To precisely determine the orientation of the spin rotation plane, the polarisation matrices of several other reflections were also considered: $(3,4/3,0)$, $(5,4/3,0)$, $(1,8/3,0)$ for all three twins. A computer code was written to fit these matrices to both sinusoidal structures with arbitrary spin direction and helical structures with arbitrary spin rotation plane. Spherical coordinates $(\theta,\varphi)$ were used to parametrize the spin direction in the case of sinusoidal structures, or the vector $\bf n$ normal to the rotation plane in the case of helical structures (where the $c$ axis corresponds to $\theta=0^\circ$). The equally populated S-domains were generated from the magnetic structure by applying one of the mirror planes ($m_{yz}$ and $m_{xz}$) or the glide plane ($n_{xy}(1/2,1/2,0)$) and the contribution from all domains were averaged. The other fitted parameters were the beam polarisation ($\mathcal{P}_\text{xx}$) and the ellipticity of the spin helix (ratio of the moment length along the two main axis of the ellipsoid, where $\mu=m_\mathbf{v}/m_\mathbf{u}$, $\mathbf{u}=\mathbf{n}\times (\mathbf{c}\times\mathbf{n})$ and $\mathbf{v}=\mathbf{u}\times\mathbf{n}$). Fitting was performed by minimizing the function $R_\text{w}^2=\sum w|\mathcal{P}_\text{OBS}-\mathcal{P}_\text{CALC}|^2/\sum w\mathcal{P}_\text{OBS}^2$, where $w=1/\sigma^2$ and $\sigma$ is the counting error and the summation was over all elements of all measured polarisation matrices.

By fitting this general model to all the measured polarisation matrices the magnetic structure was found to be helical with spins rotating in the $ac$ plane; the resulting polarisation matrices are listed in the last three rows of Table \ref{tab:polmeas}. The extracted parameters are: $\mathcal{P}_\text{xx}=0.887$, $\theta=87^\circ$, $\varphi=91^\circ$, $m_a/m_c=0.99$. With these parameters excellent agreement with the data could be achieved ($R_\text{w}=0.074$). To check how sensitive the refinement is to the spin-plane orientation, the values of $\theta$ and $\varphi$ were fixed to other directions. All other possible structures ($ab$ and $bc$ helical, $a$, $b$ and $c$ sinusoidal) give a value of $R_\text{w}$ that is more than four times larger. According to the result the spin components along $a$ and $c$ are close to equal in agreement with Ref.\ \onlinecite{Chapon2011}, whereas the single crystal and powder refinements show larger $a$ component by a factor of 1.2. 

Using spherical neutron polarimetry it has been possible to prove unambiguously that $\alpha$-CaCr$_2$O$_4$  has a $\sim120^{\circ}$ helical magnetic structure with spin moments rotating in the $ac$ plane perpendicular to the ordering wavevector ${\bf k}=(0,\sim1/3,0)$. The magnetic structure is illustrated in Fig.\ \ref{fig:sstructure}.


\section{Discussion}

The $120^{\circ}$ helical magnetic structure of $\alpha$-CaCr$_2$O$_4$ is exactly the structure expected for an ideal triangular antiferromagnet with Heisenberg nearest neighbor interactions that are all equal. This is a surprising result given that the triangular plane is in fact distorted with two inequivalent Cr$^{3+}$ ions and four independent nearest neighbor exchange interactions. Other distorted triangular antiferromagnets that have been investigated so far all show departures from ideal triangular behaviour with ordering wavevectors that deviate from $(0,1/3,0)$ e.g.\ Cs$_{2}$CuCl$_{4}$ (where ${\bf k}=(0,0.472,0)$), NaMnO$_{2}$ (where ${\bf k}=(1/2,1/2,0)$) and CuCrO$_{2}$ (where ${\bf k}=(0,0.329,0)$). In this section we solve this apparent contradiction by investigating how the ordering wavevector and relative orientation of the spin moments in $\alpha$-CaCr$_2$O$_4$ vary as a function of the exchange interactions.

The possible interactions between the magnetic moments can be deduced from the crystal structure. Figure \ref{fig:dist} shows the location of the Cr$^{3+}$ ions on the triangular plane. There are four different nearest neighbour Cr$^{3+}$--Cr$^{3+}$ distances, $d_\text{ch1}$, $d_\text{ch2}$, $d_\text{zz1}$ and $d_\text{zz2}$, ranging from 2.882 to 2.932 \AA\ at 2.1 K which are identified by the different colored lines. The Cr$^{3+}$ ions are surrounded by oxygen octahedra which lower the energy of the $t_\text{2g}$ orbitals and since there are three electrons in the 3d shell, each of the $t_\text{2g}$ orbitals is occupied by one single electron. Neighboring CrO$_6$ octahedra are edge-sharing and this along with the short distances between the magnetic ions implies that the exchange interactions are dominated by direct overlap of singly occupied $t_\text{2g}$ orbitals and are therefore antiferromagnetic. \cite{Goodenough1960} Since direct exchange interactions are highly sensitive to ionic separation we expect four different exchange constants, $J_\text{ch1}$, $J_\text{ch2}$, $J_\text{zz1}$ and $J_\text{zz2}$, corresponding to the four nearest-neighbor distances as labeled in Fig.\ \ref{fig:dist}. These interactions are expected to be Heisenberg with no anisotropy because the occupied $t_\text{2g}$ multiplet has quenched orbital angular momentum. Although there are two triangular layers per unit cell, they are in fact the same due to the the reflection plane $m(1/4,y,z)$ of the $Pmmn$ space group. The interlayer interactions are also expected to be antiferromagnetic but much weaker and to occur via superexchange through intermediate oxygen ions. Furthermore these interactions are unfrustrated and do not influence the ordering $\bf k$ wavevector. Next nearest neighbor intraplane interactions may be present and would influence $\bf k$, as reported for CuCrO$_2$, \cite{Poienar2010} another member of the delafossite arisotype which has similar Cr$^{3+}$--Cr$^{3+}$ distances. 
\begin{figure}[!htb]
  \centering
  \includegraphics[width= 86 mm]{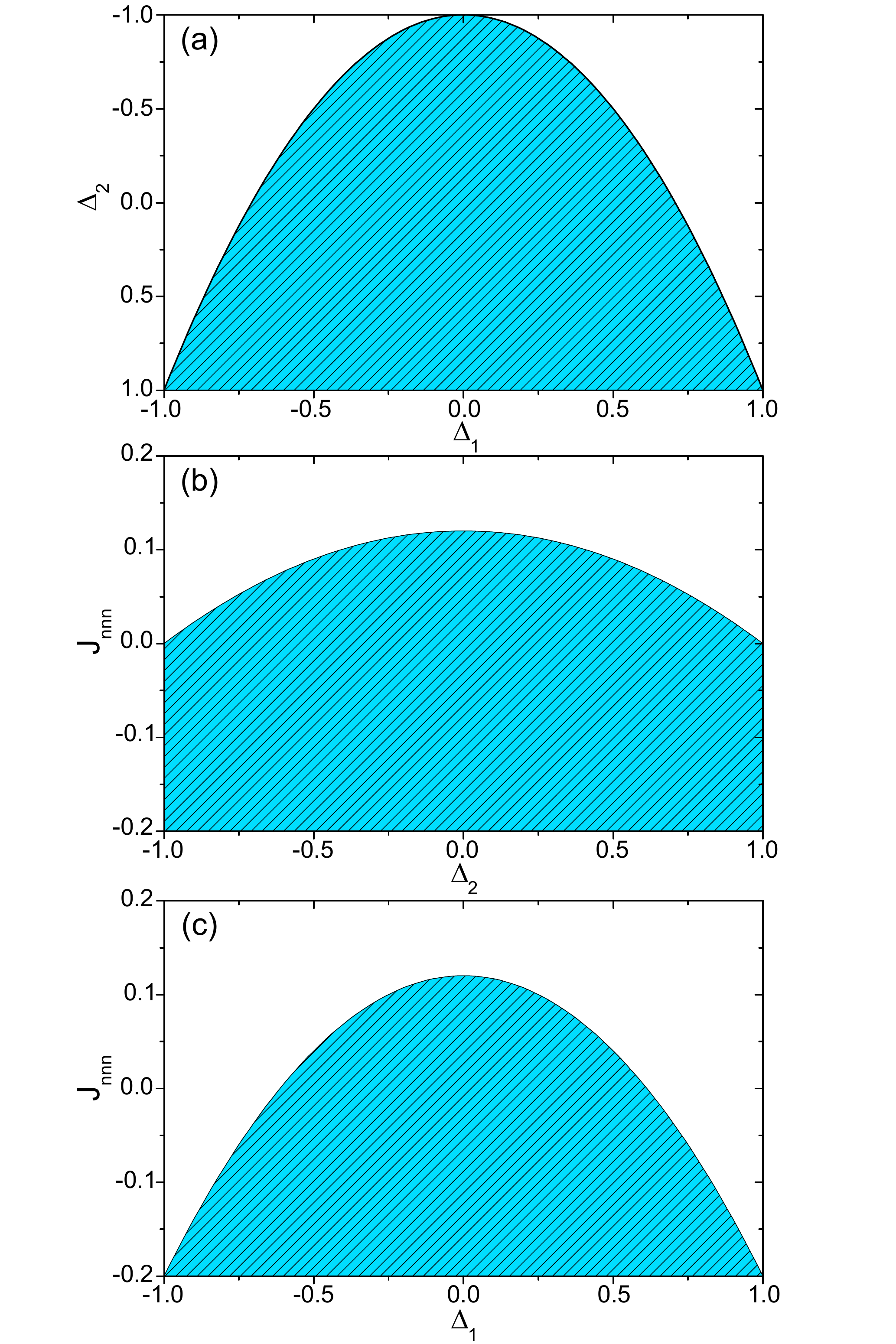}
  \caption{Magnetic phase diagrams as a function of $\Delta_1$, $\Delta_2$, and $J_\text{nnn}$ with $\Delta_3$ fixed to zero for all graphs. Inside the blue region ${\bf k}=(0,1/3,0)$ with $120^\circ$ spin structure ($\varphi^{1a}=0$, $\varphi^{1b}=240^{\circ}$, $\varphi^{2a}=120^{\circ}$ and $\varphi^{2b}=0^{\circ}$). Elsewhere ${\bf k}=(0,1,k_z)$ and the spin angles vary.}
  \label{fig:J}
\end{figure}

The magnetic Hamiltonian of $\alpha$-CaCr$_2$O$_4$ can be expressed as the sum of a Hamiltonian containing nearest-neighbor interactions and a Hamiltonian containing next-nearest-neighbor interactions, while the much weaker interlayer interactions have been ignored
\begin{align}
\mathcal{H}=\mathcal{H}_\text{nn}+\mathcal{H}_\text{nnn}.  \nonumber
\end{align}
The nearest neighbor Hamiltonian has the form
\begin{align}
\mathcal{H}_\text{nn} =\displaystyle\sum_\text{nm}&J_\text{ch1}{\bf S}^{2a}_\text{nm}\cdot\left({\bf S}^{2b}_\text{nm}+{\bf S}^{2b}_\text{(n-1)m} \right)  \nonumber\\
+&J_\text{ch2}{\bf S}^{1a}_\text{nm}\cdot\left({\bf S}^{1b}_\text{nm}+{\bf S}^{1b}_\text{(n-1)m}\right)    \nonumber\\
+&J_\text{zz1}{\bf S}^{2a}_\text{nm}\cdot\left({\bf S}^{1a}_\text{n(m-1)}+{\bf S}^{1b}_\text{n(m-1)}\right)\nonumber\\                      +&J_\text{zz1}{\bf S}^{2b}_\text{nm}\cdot\left({\bf S}^{1a}_\text{(n+1)m}+{\bf S}^{1b}_\text{nm}\right)    \nonumber\\
+&J_\text{zz2}{\bf S}^{2a}_\text{nm}\cdot\left({\bf S}^{1a}_\text{nm}+{\bf S}^{1b}_\text{nm}\right)        \nonumber\\
+&J_\text{zz2}{\bf S}^{2b}_\text{nm}\cdot\left({\bf S}^{1a}_\text{(n+1)(m-1)}+{\bf S}^{1b}_\text{n(m-1)}\right),
\end{align}
where ${\bf S}^{i}_\text{nm}$ is the spin of Cr$^{3+}$ site (i$\in\{1a, 1b, 2a, 2b\}$), and $n$ and $m$ are the indices of the unit cell along the $b$ and $c$ axes respectively. The next-nearest-neighbor Hamiltonian has the form
\begin{align}
\mathcal{H}_\text{nnn} =J_\text{nnn}\displaystyle\sum_\text{nm}
&{\bf S}^{2a}_\text{nm}\cdot\left({\bf S}^{2a}_\text{(n+1)m}+{\bf S}^{1a}_\text{(n+1)m}+{\bf S}^{1a}_\text{(n+1)(m-1)}\right) \nonumber\\
+&{\bf S}^{2b}_\text{nm}\cdot\left({\bf S}^{2b}_\text{(n+1)m}+{\bf S}^{1b}_\text{(n+1)m}+{\bf S}^{1b}_\text{(n+1)(m-1)}\right) \nonumber\\
+&{\bf S}^{1a}_\text{nm}\cdot\left({\bf S}^{1a}_\text{(n+1)m}+{\bf S}^{2b}_\text{n(m+1)}+{\bf S}^{2b}_\text{nm}        \right) \nonumber\\
+&{\bf S}^{1b}_\text{nm}\cdot\left({\bf S}^{1b}_\text{(n+1)m}+{\bf S}^{2a}_\text{(n+1)m}+{\bf S}^{2a}_\text{(n+1)(m+1)}\right),
\end{align}
where a single $J_\text{nnn}$ exchange constant is assumed for all next-nearest-neighbor interactions. 
 
The Hamiltonian can be used to explore the magnetic structure of $\alpha$-CaCr$_2$O$_4$ as a function of the exchange interactions. In order to do this all the spins are treated as classical vectors of the same length and are restricted to be coplanar. The aim is to understand how the in-plane $\bf k$-vector (both $k_y$ and $k_z$ components) and the directions of the spin moments $\varphi^i_\text{nm}$ (for the $i$th spin of the ($n,m$)-th unit cell) depend on the magnetic interactions. $\varphi^i_\text{nm}$ can be written in terms of $\varphi^i$, the angle of spin ${\bf S}^i$ in the first unit cell, and the ordering vector: $\varphi^i_\text{nm}=\varphi^i+{\bf k}\cdot {\bf r}_\text{nm}$, where ${\bf r}_\text{nm}$ is the position of the unit cell. The Hamiltonian can be re-expressed in terms of $\bf k$ and $\varphi^i$ by rewriting the dot product of the spins, for example
\begin{align}
{\bf S}^i_\text{nm}\cdot{\bf S}^{i'}_\text{n'm'} = S^2 \cos \left[ \left( \varphi^{i'} - \varphi^{i} \right) + {\bf k} \cdot \left( {\bf r}_\text{n'm'} - {\bf r}_\text{nm} \right) \right].
\end{align}
$\varphi^{1a}$ can be set to zero due to the $O(3)$ invariance of the Heisenberg Hamiltonian. 

A linear transformation was performed on the exchange constants, which was found to give more insight into the results as will be explained later
\begin{align}
  J_\text{mean}&=\frac{1}{4}(J_\text{ch1}+J_\text{ch2}+J_\text{zz1}+J_\text{zz2}),                  \nonumber\\
  \Delta_1 &=\frac{1}{2J_\text{mean}} (J_\text{zz1}-J_\text{zz2}),                        \nonumber\\
  \Delta_2 &=\frac{1}{2J_\text{mean}} (J_\text{ch1}-J_\text{ch2}),                        \nonumber\\
  \Delta_3 &=\frac{1}{4J_\text{mean}} [(J_\text{ch1}+J_\text{ch2})-(J_\text{zz1}+J_\text{zz2})].
\end{align}
$J_\text{mean}$ gives the overall energy scale and does not affect the zero temperature ground state structure, thus together with $J_\text{nnn}$ only four important parameters remain. Finally, to determine the ground state values of $\bf k$, $\varphi^{1b}$, $\varphi^{2a}$ and $\varphi^{2b}$ for a given set of exchange interactions $\Delta_1$, $\Delta_3$, $\Delta_3$ and $J_\text{nnn}$ the energy of the Hamiltonian was minimized by performing a constrained linear optimization using a simplex minimisation algorithm.

Figure \ref{fig:J} shows the in-plane $\bf k$-vector as a function of $\Delta_1$, $\Delta_2$ and $J_\text{nnn}$ with $\Delta_3$ fixed to zero. 
The reason for the linear transformation of the exchange constants is clear from these plots. 
They show that the magnetic structure is commensurate (${\bf k}=(0,1/3,0)$), with a $120^\circ$ ground state ($\varphi^{1a}=0$, $\varphi^{1b}=240^{\circ}$, $\varphi^{2a}=120^{\circ}$ and $\varphi^{2b}=0^{\circ}$ over a large volume of $\Delta_1$, $\Delta_2$ and $J_\text{nnn}$ parameter space. Although the crystal structure is distorted, the nature of the distortion is such that when $\Delta_3$ is constrained to zero, commensurate $120^\circ$ magnetic order is stabilized unless the magnitude of $\Delta_1$, $\Delta_2$ and $J_\text{nnn}$ are much greater than 0. 
\begin{figure}[!htb]
  \centering
  \includegraphics[width= 86 mm]{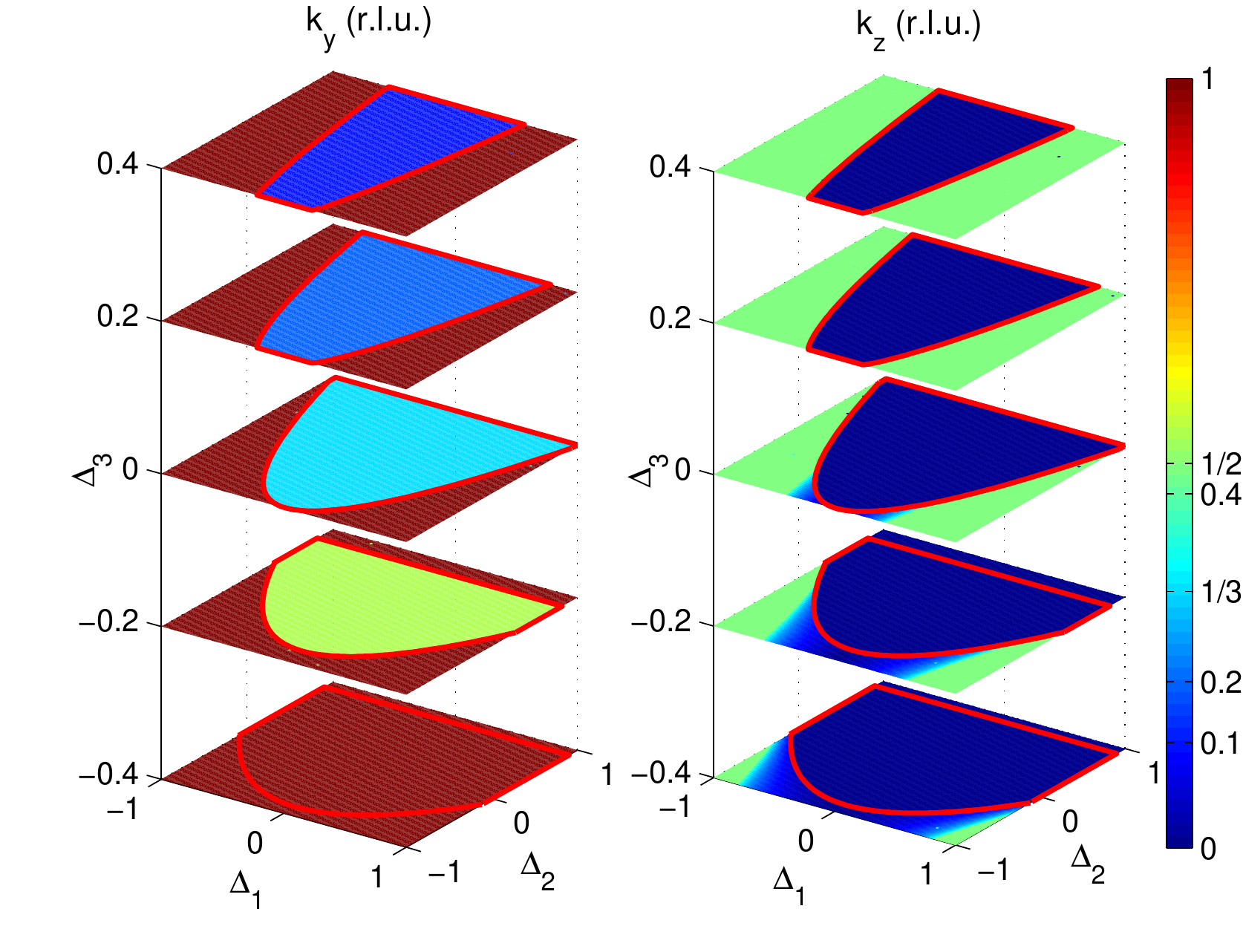}
  \caption{Magnetic phase diagrams showing the dependence of $\bf k$ on $\Delta_1$, $\Delta_2$ and $\Delta_3$; (a) displays $k_y$ and (b) $k_z$ in units of $2\pi$. For $\Delta_3=0$, ${\bf k}=(0,1/3,0)$ and the spins form a $120^{\circ}$ spin structure ($\varphi^{1a}=0$, $\varphi^{1b}=240^{\circ}$, $\varphi^{2a}=120^{\circ}$ and $\varphi^{2b}=0^{\circ}$) within the red border. For $\Delta_3\neq0$, $\bf k$ becomes incommensurate.}
  \label{fig:Jlim}
\end{figure}

\begin{figure}[!htb]
  \centering
  \leavevmode
  \includegraphics[width= 86 mm]{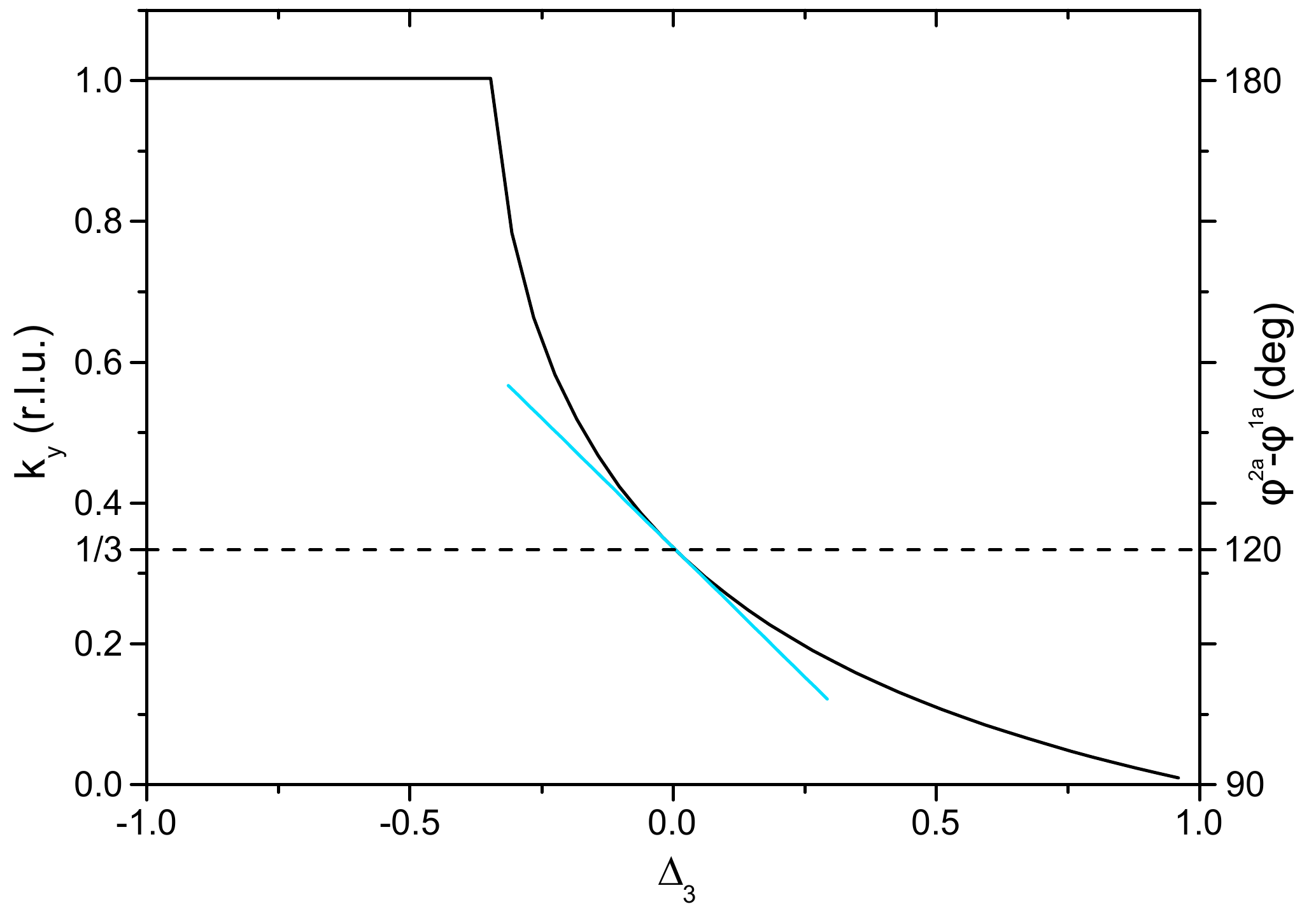}
  \caption{Dependence of the y component of $\bf k$ on $\Delta_3$ ($\Delta_1=\Delta_2=J_\text{nnn}=0$), right scale shows the phase difference between ${\bf S}^{1a}$ and ${\bf S}^{2a}$. At $\Delta_3=0$ the tangent of the slope (blue line) is -0.74.}
  \label{fig:d3ky}
\end{figure}

If $\Delta_3$ is varied as well the $\bf k$-vector changes continuously, as shown in Fig.\ \ref{fig:Jlim}. Within the red borders on the plots, all four $\varphi^i$ angles are determined only by the $\bf k$-vector, according to the expression $\varphi^i=({\bf k}+(0,1,0))\cdot{\bf \rho}^\text{i}$ where ${\bf \rho}^\text{i}$ is the position of ${\bf S}^\text{i}$ in the first unit cell. Outside these regions the ordering wavevector locks into a new set of value (${\bf k}=(0,1,k_\text{z})$) and the relative phases inside the unit cell are independent of $\bf k$. Figure \ref{fig:d3ky} shows how {\bf k} varies with $\Delta_3$, (all other parameters are fixed to zero). Even a slight deviation of $\Delta_3$ from zero drives $\bf k$ incommensurate. The slope of $k_\text{y}$ at the point $\Delta_3=0$ is $-0.74$, this can be used to estimate the value of $\Delta_3$ for $\alpha$-CaCr$_2$O$_4$. From powder diffraction data the upper limit for the difference between $k_\text{y}$ and the commensurate 1/3 value is d$(k_\text{y})<0.004$. This means, $|\Delta_3|<|1/(-0.74)| $d$(k_\text{y})\sim 0.005$. So the only condition for the observed magnetic structure is that $|\Delta_3|<0.005$. Comparison of the different Cr$^{3+}$--Cr$^{3+}$ nearest neighbor distances reveal that the average of $d_\text{zz1}$ and $d_\text{zz2}$ (2.907 \AA) is almost the same as the average of $d_\text{ch1}$ and $d_\text{ch2}$ (2.912 \AA). This explains why the direct exchange interactions appear to obey $(J_\text{zz1}+ J_\text{zz2})\approx(J_\text{ch1}+ J_\text{ch2})$ giving $|\Delta_3|\approx 0$.

Although $\alpha$-CaCr$_2$O$_4$ is distorted from ideal triangular symmetry with two inequivalent Cr$^{3+}$ ions and four different exchange interactions, the nature of the distortion is such that the average of the exchange interactions along any direction are approximately equal. On larger length scales the Hamiltonian is in fact spatially isotropic providing an explanation for the observed $\sim120^\circ$ helical magnetic ordering.


\section{Conclusion}

In this paper we presented a detailed investigation of the magnetic order in the distorted triangular antiferromagnet $\alpha$-CaCr$_2$O$_4$. The first single-crystal growth is reported and bulk properties measurements along with powder and single-crystal X-ray and neutron diffraction were performed. The ordering wavevector was found to be close to commensurate ${\bf k}=(0,1/3,0)$ and we unambiguously determined that the magnetic ordering is helical with the spin moments forming a $\sim120^\circ$ structure between nearest neighbors in the $ac$ plane. It is important to note that a small deviation from the commensurate wavevector is probable due to the orthorhombic crystal symmetry. The apparent contradiction between the observed ideal magnetic structure and the distorted crystal structure was resolved by exploring the magnetic phase diagram. The distortions in the triangular plane are only significant on individual nearest neighbor distances, while on the length scale of a unit cell the average exchange interactions along any triangular axis are approximately equal.

\begin{acknowledgments}
We acknowledge P. J. Brown for valuable advice on the spherical polarimetry analysis. The work was partially performed at SINQ, Paul Scherrer Institute, Villigen, Switzerland.
\end{acknowledgments}

\bibliography{alphaCaCr2O4_arXiv}

\end{document}